\documentclass[preprint,10pt]{sigplanconf}
\pdfoutput=1

\usepackage{pbox}
\usepackage{amsfonts}
\usepackage{amssymb}
\usepackage{color}
\usepackage{amsmath}

\usepackage{amsthm}
\usepackage{url}
\usepackage{stmaryrd}   
\usepackage{graphicx}   
\usepackage{enumitem}
\usepackage{datetime}   
\usepackage{wrapfig}
\usepackage{listings}
\usepackage{courier}

\usepackage[draft]{hyperref}

\newcommand{\ignore}[1]{}

\newcommand{\dataset}{\mathcal{D}}

\newcommand{\corpus}{\mathcal{C}}

\newcommand{\neuron}{\mathcal{N}}
\newcommand{\sigmoid}{\mathcal{S}}

\newcommand{\states}{\mathcal{Q}}
\newcommand{\rnn}{\mathcal{R}}

\newcommand{\algosample}{\textsc{Sample}}

\newcommand{\trans}{\delta}
\newcommand{\transition}[4]{#1~\xrightarrow{#3,#4}~#2}

\newcommand{\kld}[2]{\mbox{$D_{KL}(#1 \| #2)$}}

\newcommand{\Dir}{\mbox{Dir}}
\newcommand{\Categorical}{\mbox{Categorical}}

\newcommand{\Obsalphabet}{\Sigma}
\newcommand{\Absalphabet}{\widehat{\Obsalphabet}}
\newcommand{\obssymbol}{s}
\newcommand{\abssymbol}{\widehat{\obssymbol}}

\newcommand{\eval}[2]{\llbracket {#1} \rrbracket_{#2}}

\newcommand{\Assign}{\mbox{\textsf{:=}}}

\newcommand{\seclabel}[1]{\label{sec:#1}}
\newcommand{\secref}[1]{Section~\ref{sec:#1}}

\newcommand{\salento}{{\sc Salento}}
\begin{document}

\title{Finding Likely Errors with Bayesian Specifications
\thanks{This research was supported by DARPA MUSE award \#FA8750-14-2-0270. The
views, opinions, and/or findings contained in this article are those of the
authors and should not be interpreted as representing the official views or
policies of the Department of Defense or the U.S. Government. }
}



\authorinfo{Vijayaraghavan Murali}
{Rice University}
{vijay@rice.edu}

\authorinfo{Swarat Chaudhuri}
{Rice University}
{swarat@rice.edu}

\authorinfo{Chris Jermaine}
{Rice University}
{cmj4@rice.edu}





\maketitle

\begin{abstract}
We present a Bayesian framework for learning probabilistic
specifications from large, unstructured code corpora, and a method
to use this framework to statically detect {\em anomalous}, hence likely buggy, program behavior.
The distinctive insight here is to build a statistical
model that correlates {\em all specifications} hidden inside a corpus
with the syntax and observed behavior of programs that implement these
specifications. During the analysis of a particular program, this
model is conditioned into a posterior distribution that prioritizes
specifications that are relevant to this program. This allows accurate
program analysis even if the corpus is highly heterogeneous. The
problem of finding anomalies is now framed quantitatively, as a problem of computing a
distance between a ``reference distribution'' over program
behaviors that our model {\em expects} from the
program, and the distribution over behaviors that the program actually
produces.

We present a concrete embodiment of our framework that combines a
topic model and a neural network model to learn specifications, and
queries the learned models to compute anomaly scores.  We
evaluate this implementation on the task of detecting anomalous usage
of Android APIs. Our encouraging experimental results show that the
method can automatically discover subtle errors in Android
applications in the wild, and has high precision and recall compared
to competing probabilistic approaches.

\end{abstract}

\section{Introduction}\seclabel{intro}

Over the years, research on automated bug finding in programs has had
many real-world
successes~\cite{Bessey:2010:FBL:1646353.1646374,Ball:2011:DSM:1965724.1965743}.
However, one perennial source of difficulty here is the need for
formal specifications.  Traditional approaches in this area require
the user to specify correctness properties; any property that is not
specified is outside the scope of reasoning. However, formally
specifying real-world software is a difficult task that users often
refuse to undertake.

A natural response to this difficulty is to {\em automatically learn}
specifications of popular software components like APIs and
frameworks. The availability of large corpora of open-source code ---
{\em Big Code}~\cite{raychev2015predicting}, in the language of some
recent efforts --- makes this idea especially appealing. By analyzing
these corpora, one can generate numerous examples of how
real-world programs use a set of components. Statistical methods can
then be used to learn common patterns in these examples. According to the
well-known thesis that bugs are {\em anomalous} behaviors~\cite{bugsdeviant,hangal2002tracking}, 
a program whose use of
the components significantly deviates from these ``typical'' usage
patterns can be flagged as erroneous.

The problem of specification learning has been studied for a long
time~\cite{ABL02,Ammons03,Alur05,GouesTACAS09}.  However, existing approaches to the problem face two
basic issues when applied to large code corpora.  First,
examples derived from such a corpus can be {\em noisy}. While programs
in a mature corpus are likely to be correct on the average, not
all examples extracted from such a corpus represent correct
behavior. Second, such a corpus is fundamentally {\em heterogeneous},
and may contain many different specifications, some of them mutually
contradictory. For example, it may be legitimate to use a set of APIs
in many different ways depending on the context, and a large enough
corpus would contain instances of all these usage patterns. A specification
learning tool should distinguish between these patterns, and a
bug-finding tool should only compare a program with the patterns that
are relevant to it.

Among existing methods for specification learning, the majority  
follow a traditional, qualitative view of program correctness. In this
view, a specification is a set of program behaviors (e.g., sequences
of calls to API methods), and a behavior is either correct (in the
specification) or incorrect (outside the specification). Such an
approach is not robust to noise because its belief in the correctness
of a behavior does not change smoothly with the behavior's
observed frequency. A small number of incorrect examples can 
persuade the method that the behavior is fully correct.

An obvious fix to this problem is to view a specification as a {\em
probabilistic} rather than a boolean model. Such a specification
assigns quantitative {\em likelihood} values to observed program behaviors, with
higher likelihood representing greater confidence in a behavior's
correctness. Some recent work adopts this view by modeling
program behaviors using models like $n$-grams~\cite{nguyenICSE15,bugramASE16} and
recurrent neural networks~\cite{vechevPLDI14}. To find bugs using such a model, one
generates behaviors of the target program using static or dynamic 
analysis, then evaluates the likelihood of these
behaviors~\cite{WangCMT16}.

While robust to noise, approaches of this sort have a basic difficulty
with heterogeneity. The root of this difficulty is that these methods learn a
{\em single} probability distribution over program behaviors. For
example, if $\Psi_1$ and $\Psi_2$ are two common but distinct patterns
in which programs in a corpus use a set of APIs, these approaches
would learn a specification that is a {\em mixture} of $\Psi_1$ and
$\Psi_2$. During program analysis, such a mixture would assign low or
meaningless likelihoods to behaviors that match {one of}, but not
both, $\Psi_1$ and $\Psi_2$. As behaviors from a given program are
likely to follow only one of the two patterns, this phenomenon would
lead to inaccurate analysis.

In this paper, we present a {\em Bayesian} approach to specification
learning and bug finding that is robust to 
heterogeneity and noise. Our key insight is to build a ``big'' statistical model
that captures {\em the entire gamut of specifications} in
an unstructured code corpus. More precisely, our model
learns a joint probability distribution that relates hidden
specifications $\Psi$ with {\em syntactic features} $X$ that describe
what implementations of these specifications ``look like''. When using
this model to analyze a particular program $F$, we {specialize} it
into a {\em posterior distribution} $P(\Psi | X_F)$ over
specifications, conditioned on the features $X_F$ of $F$. Intuitively,
this distribution assigns higher weight to specifications for
programs that ``look like'' $F$, and can be seen as the part of the
model that is relevant to $F$.

This model architecture can tolerate high (in principle, unbounded)
amounts of heterogeneity in the corpus. Suppose that the programs in
the corpus use a set of APIs following distinct patterns
$\Psi_1,\dots, \Psi_n$, but that programs that look like $F$ (i.e.,
have feature set $X_F$) tend to follow $\Psi_1$. During training, our
framework learns this correlation between $X_F$ and
$\Psi_1$. This means that the posterior distribution $P(\Psi | X_F)$
puts a high weight on $\Psi_1$ and low weights on
$\Psi_2,\dots, \Psi_n$, and that effectively,
correctness analysis of $F$ happens with respect to $\Psi_1$
rather than any other specification.

Our second key idea is to frame the detection of likely
errors as an operation over probability distributions.  We assume, for
each program $F$, a distribution $P_F(\theta)$ over the behaviors of
the program. This distribution --- a {\em probabilistic behavior
  model} --- may be learned from data, or, as is the
case in this paper, be a {\em definition} that is a parameter of the
framework.
This allows us to develop a model
$P(\theta |\Psi)$ of the behaviors $\theta$ of a program that follows
a specification $\Psi$. When combined with the posterior distribution
$P(\Psi | X_F)$ for $\Psi$, this model gives us a ``reference
distribution'' $P(\theta | X_F)$ over behaviors that the model {\em
  expects} from a program that looks like $F$. The {\em anomaly score}
of $F$, which quantifies the extent to which $F$ behaves abnormally,
is now defined as a {statistical distance} 
(in particular, the Kullback-Leibler  divergence~\cite{KLD}) 
between  $P(\theta | X_F)$ and $P_F(\theta)$.

Our Bayesian approach is a framework, meaning that it can be
implemented using a wide range of concrete statistical models.  The
particular instantiation we present in this paper is a combination of the
popular topic model known as Latent Dirichlet Allocation
(LDA)~\cite{LDA}, and a class of neural networks that are {\em
  conditioned} on a topic model~\cite{topicRNN}. To compute the anomaly
score for a program $F$, 
\ignore{we use a quantitative program analysis that
repeatedly queries this model for the likelihood of different
behaviors of $F$, then uses a meet-over-all-paths computation to
aggregate these likelihood values into an estimate of the anomaly
score.}
we repeatedly query this model for the likelihood of different behaviors
of $F$, and then aggregate these likelihood values into an estimate of
the anomaly score.

We evaluate our implementation in the task of finding erroneous API
usage in Android applications. Using three APIs as benchmarks,
we show that the tool can automatically discover subtle API bugs in
Android applications in the wild. These violations range from GUI bugs
to inadequate encryption strength. Some of these errors are difficult
to characterize in logic-based specification notations,
indicating the promise of our approach in settings where traditional
formal methods are hard to apply. We also demonstrate that the method
has good precision recall and is more robust to heterogeneity than
a comparable non-Bayesian approach.

Now we summarize the contributions of this paper:
\begin{itemize}[leftmargin=*]
\item We present a novel Bayesian framework for learning 
specifications from large code corpora. 
(\secref{formulation})

\item We offer a novel formulation of the problem of finding
  anomalous program behavior as the problem of computing a 
  distance between a program and a reference distribution. (\secref{formulation})

\item We present an instantiation our framework with
  a topic model and a topic-conditioned recurrent neural network.
(\secref{instantiation})

\item We evaluate the approach on the problem of detecting anomalous API
  usage in a suite of Android applications (\secref{eval})
\end{itemize}



\section{Overview}\seclabel{overview}

In this section, we present an overview of our approach, with the help
of an illustrative example.

\subsection{Modeling Framework and Workflow}

Our approach has the following key aspects.  First, we assume the
existence of a {\em specification} $\Psi$ for each program
$F$. However, unlike traditional approaches that {\em start} with a
formal specification, $\Psi$ in our context is not
observable. Instead, what is observable is $X_F$, a set of syntactic
\emph{features} for $F$.  The features are evidence, or data, that
inform our opinion as to the unseen specification $\Psi$.  
In Bayesian fashion, our uncertainty about $\Psi$ is 
formalized as a {\em posterior distribution} $P(\Psi | X_F)$, which
measures the extent to which we believe that $\Psi$ is the correct
specification for $F$, given the evidence.

Second, our framework allows for uncertainty regarding the {\em
  behaviors} $\theta$ --- defined as sequences of observable actions
--- that a given program $F$ produces. This uncertainty comes from the
fact that we do not exactly know the inputs on which the program will
run, and is captured by a probability distribution
$P_F(\theta)$.  The framework also allows for a distribution
$P(\theta|\Psi)$ over the behaviors of programs that implement a given
specification $\Psi$.  This uncertainty can come from the fact that we
do not know the inputs to implementations of $\Psi$, or the fact that we
may have never seen a specification exactly like $\Psi$ before, so
that we have to guess the behavior of a program implementing $\Psi$.

Our \emph{a priori} belief about the relationships between
specifications and the features and behaviors of their implementations
is given by a joint distribution $P (\theta, X_F, \Psi)$. Our third
key idea is that this distribution is informed by
data extracted from a corpus of code.  This information is taken into
account formally during a {\em learning phase} that fits the joint
distribution prior model to the data.

Finally, in the {\em inference phase}, we frame bug detection as a problem of computing a
quantitative {\em anomaly score}.  In traditional correctness
analysis, the semantics of programs and specifications are given
by sets, and one checks if the {\em set difference} between a program and
a specification is empty. Our formulation is a quantitative
generalization of this, and defines the anomaly score for a program
$F$ as the {\em Kullback-Leibler (KL) divergence}~\cite{KLD}  between
the behavior distribution $P_F(\theta)$ for $F$, and the posterior
distribution $P(\theta | X_F)$ that the model expects from
$F$. Correctness analysis amounts to checking whether this score
is below a threshold.

\begin{figure}
\hspace{-0.5cm}
\includegraphics[scale=0.35]{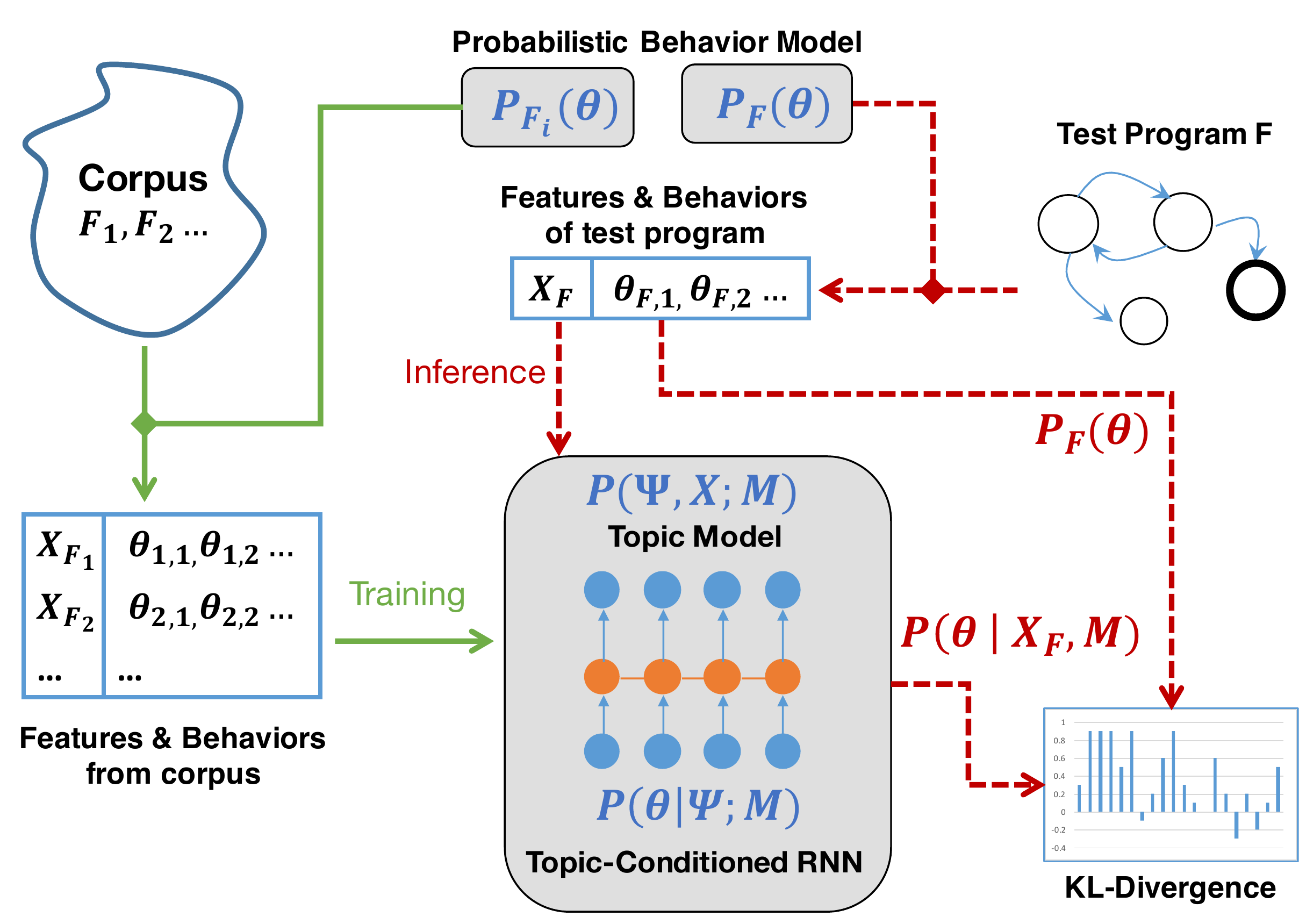}
\caption{Workflow, with instantiations in grey boxes}
\label{fig:workflow}
\vspace{-0.2in}
\end{figure}

\begin{figure*}
\begin{tabular}{ccc}
\begin{minipage}{0.2\textwidth}
\includegraphics[scale=0.25]{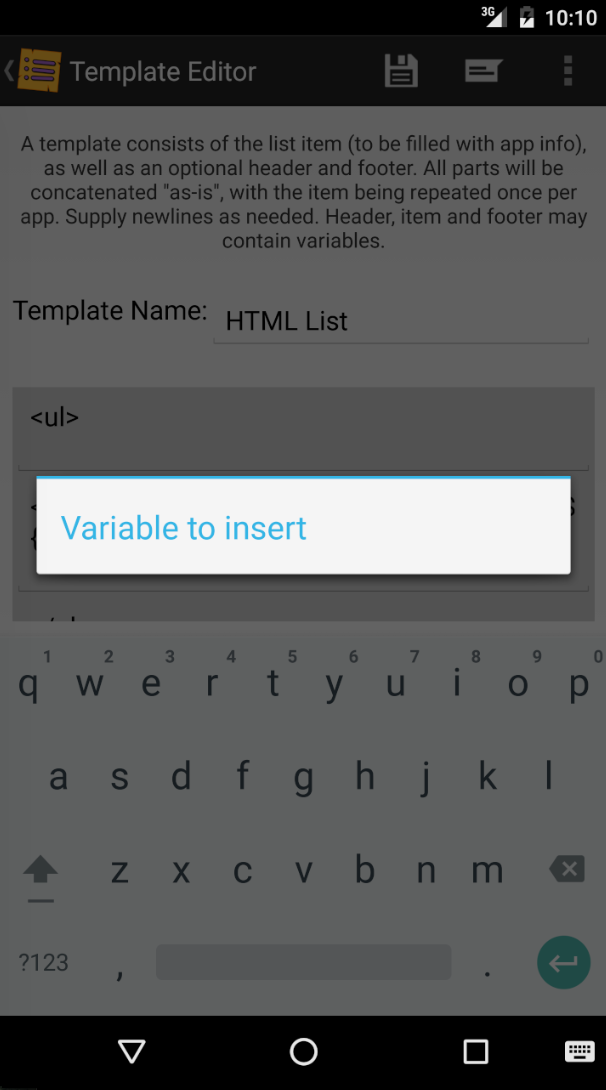}
\end{minipage}
&
\begin{minipage}{0.25\textwidth}
\includegraphics[scale=0.12]{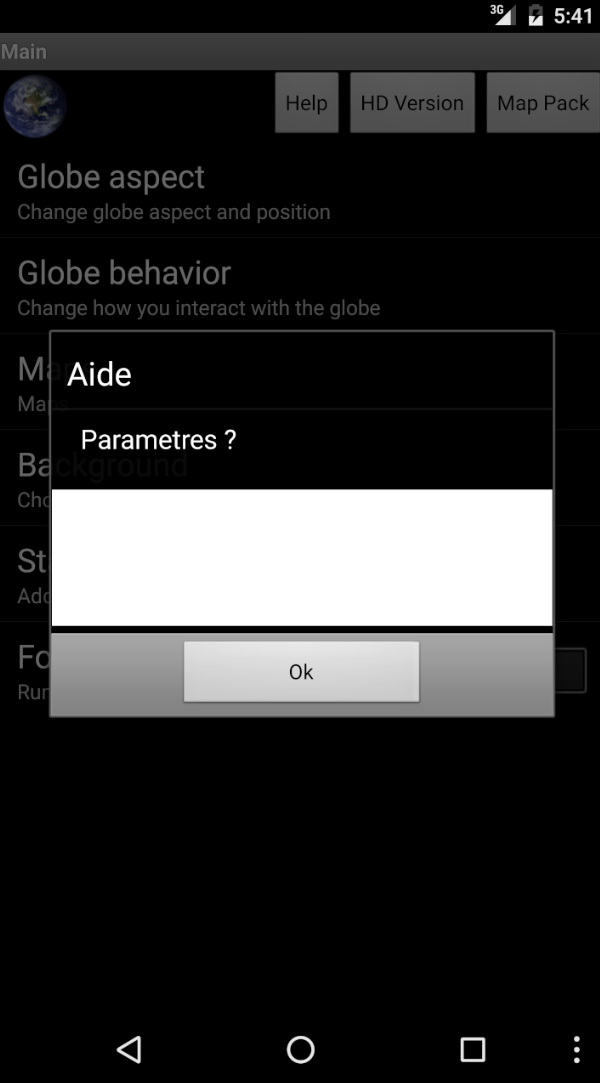}
\end{minipage}
&
\begin{tabular}{l}
\begin{lstlisting}[showlines=true,basicstyle=\fontsize{7}{7}\selectfont\ttfamily]
AlertDialog.Builder b = new AlertDialog.Builder(this);
b.setTitle(R.string.title_variable_to_insert);
if (focus.getId() == R.id.tmpl_item)
    b.setItems(R.array.templatebodyvars, this);
else if (focus.getId() == R.id.tmpl_footer)
    b.setItems(R.array.templateheaderfootervars, this);
b.show(); 
\end{lstlisting}
\\ \multicolumn{1}{c}{(b)(i)} \\
\begin{lstlisting}[showlines=true,mathescape,basicstyle=\fontsize{7}{7}\selectfont\ttfamily]
AlertDialog.Builder b = new AlertDialog.Builder(this);
b.setMessage("Parametres?");
b.setCancelable(false);
b.setView(dlgLayout);
b.setPositiveButton("Ok", new OnClickListener(){$\ldots$});
b.setTitle("Aide")
b.show();
\end{lstlisting}
\end{tabular}
\\ \hspace{-1cm}(a)(i) & \hspace{-1cm}(a)(ii) & (b)(ii)
\end{tabular}
\caption{(a) Abnormal dialog boxes discovered by our anomaly detection (b) Code
snippets corresponding to the dialog boxes}
\label{fig:ui-bug}
\end{figure*}

The workflow of our method is as in Figure~\ref{fig:workflow}.
The training and inference phases are denoted by green
(solid) edges and red (dashed) edges respectively.  During training, from
each program $F_i$ in a large corpus of programs
$F_1, F_2,\dots$, we extract a set of
syntactic features $X_{F_i}$, and sample a set of behaviors from
the distribution $P_{F_i}(\theta)$, forming the training data.
From this data, we learn the joint distribution
$P(\theta, X, \Psi ; \textbf{M})$, where {\bf M} are the model parameters.

During inference, we extract the features
$X_F$ of a given program $F$, and query the trained model 
for the distribution $P(\theta | X_F; \textbf{M})$
that tells us how $F$ {should} behave. Separately, we obtain the
distribution $P_F(\theta)$ over observed behaviors of $F$. The anomaly
score of $F$ is then computed as the KL-divergence between these
distributions.



\ignore{
We next define $P (\theta | \Psi)$, which defines the expected program
behavior given $\Psi$.  Suppose our objective is to find problematic
API usage. This means that $\theta$ could be a specific sequence of API
method calls, and $P (\theta | \Psi)$ would then measure the
likelihood that a program whose type is specified by $\Psi$ would
produce the list of calls $\theta$ on a typical input.  This model might
be, for example, a {\em topic-conditioned recurrent neural
network}~\cite{mikolov2012context} that in this case is conditioned on
the value of $\Psi$. 

After learning $\textbf{M}$ (which parameterizes the LDA model as well
as the neural network), for an observed program described by $X_F$, we can
then obtain a posterior distribution for the set of expected program
behaviors as:

$$D_1 = P(\theta | X_F, \textbf{M}) \propto \int_{\Psi} P (\theta | \Psi) P(\Psi) d\Psi$$

Intuitively, this gives as a distribution over the program behaviors that
would be expected, after analyzing the corpus, for a program whose
feature set was $X_F$. Now our objective is to estimate the
KL-divergence between $D_1$ and the program-specific distribution $D_2$
mentioned earlier.
}

\subsection{Instantiating the Framework}

An instantiation of our framework
must concretely define program features and
behaviors, and the way in which the distributions
$P(\theta, X, \Psi; \textbf{M})$ and $P_F(\theta)$ are obtained. 
In this paper, we consider a particular instantiation
where the goal is
to learn patterns in the way programs call
methods in a set of APIs.  We abstract each such call as a {\em
  symbol} from a finite set, and define a behavior $\theta$ as a
sequence of symbols.  The feature $X_F$ for a program $F$ is
the set of symbols that $F$ can generate.

A key idea in the instantiation is to capture hidden specifications
using a {\em topic model}. Here, ``topic'' is an
abstraction of the hidden semantic structure of a program. A
specification for a program $F$ is a vector of probabilities whose the
$i$-th component is the probability that $F$ follows the $i$-th
topic. For example, the topics in a given corpus may correspond to GUI
programs and bit-manipulating programs. A program that makes many
calls to GUI APIs will likely have a higher probability for the former
topic.

Specifically, we use the well-known {\em Latent
  Dirichlet Allocation (LDA)}~\cite{LDA} topic model to learn a joint distribution
$P(\Psi, X; \textbf{M})$ over the topics and features of programs. A
{\em topic-conditioned recurrent neural network} model~\cite{topicRNN},
is used to as the second model
$P(\theta | \Psi; \textbf{M})$. The joint distribution
$P(\theta, X, \Psi; \textbf{M})$ that our framework maintains can be
factored into these two distributions.

Our probabilistic model $P_F(\theta)$ for behaviors of programs $F$ is
{\em not} data-driven. This is because to learn this distribution
statistically, we would need data on the inputs that $F$ receives in
the real world.  Since such data is not available in typical code
corpora, we simply {\em assume} a definition of $P_F(\theta)$. While
many such definitions are possible, the one we pick models $F$ as a
class of automata, called {\em generative probabilistic
  automata}~\cite{ABL02,GPA1}. The distribution $P_F(\theta)$ is
simply the semantics of this automaton.


\subsection{Example}

Consider the problem of finding bugs in GUIs, where 
the right and wrong ways of invoking GUI API methods are seldom formally
defined. 
%
Specifically, consider a dialog box in a GUI that does not give the
user an option to close the box, and a dialog box that does not
display any textual content.  Clearly, such boxes violate user
expectations, and are buggy in that sense. Two such
boxes, produced by real-world Android apps, are shown in
Figure~\ref{fig:ui-bug}(a).

The code snippets responsible for these boxes are shown in
Figure~\ref{fig:ui-bug}. For example, in
Figure~\ref{fig:ui-bug}(b)(i), ${\tt b}$ is a dialog box; the method
${\tt b.setItems(...)}$ adds content to the dialog box; the method
${\tt b.show()}$ displays the box. If the branches in lines 4 and 7
are not taken, then ${\tt b.show()}$ opens the box without a ``close''
button. Note that the sequences of API calls that lead to these bugs
are not forbidden by the API, and would not be caught by a traditional
program analysis. In contrast, a statistical method like ours can
observe thousands of programs and learn that these sequences are
abnormal.

  
Operationally, to debug this program, we generated features and
behaviors from a corpus of Android apps.  Using these features, LDA
learned to classify programs by the APIs they use, and to also
distinguish between different usage patterns in the same API.
Consider the examples of dialog box creation in
Figure~\ref{fig:ui-bug}(b), where program $F_1$ in (b)(i) explicitly
specifies the items that go into the box, and the program $F_2$ in
(b)(ii) provides a ${\tt View}$ that encompasses the items that go
into the box. LDA can assign different topics to these usage
patterns. For example, the pattern used in $F_1$ could be assigned the
first topic, resulting in a topic vector ($\Psi$) $\langle 0.98, 0.01, 0.01
\rangle$, and the pattern used in $F_2$ could be assigned the second topic,
resulting in the topic vector $\langle 0.01, 0.98, 0.01 \rangle$.


Conditioned on such a topic vector $\Psi$, a topic-conditioned RNN provides the
probability of an API call sequence $\theta$, that is, $P(\theta |
\Psi)$. For instance, given the former topic vector,
a topic-conditioned RNN trained on thousands of examples
of topics and behaviors would provide a high probability to a sequence
such as: 

${\tt new~A() ~~ setTitle(...) ~~ setItems(...) ~~ show()}$

\noindent
(where ${\tt new~A()}$ is a call to the constructor) and a low
probability to an abnormal sequence such as

${\tt new~A() ~~ setTitle(...) ~~ show()}$

\noindent
as it shows a dialog without any content. However, our
probabilistic automaton model $P_{F_1}(\theta)$ of $F_1$ would assign
about 0.66 and 0.33
probability, respectively, to these sequences.
In general, the 
KL-divergence between the two distributions will be high, causing $F_1$ to be flagged as anomalous.


\section{Bayesian Specification Framework}\seclabel{formulation}

In this section, we formalize our framework, along with the
problems of specification learning and anomaly detection.

\subsection{Program Behaviors and Features}
\label{subsec:behaviors}

Our framework is parameterized by a programming language.  Each program
in the language has a syntax and an operational semantics. Because the
details of the language do not matter to the framework, we do not
concretely define this syntax and semantics. Instead, we assume that
the syntax of each program $F$ can be abstracted into a {\em feature
  set} $X_F$. For instance, such features can include syntactic
constructs, assertions, and natural language comments.
We also assume that program actions during 
execution can be abstracted into a finite alphabet $\Sigma$ of {\em
  observable symbols} (including an {\em empty symbol} $\epsilon$). We
model program executions as {\em behaviors} $\theta$, defined to be
words in $\Sigma^*$.
A behavior is the result of a probabilistic
generative process that takes place when a program is executed.
Accordingly, we assume a {\em probabilistic behavior model} of $F$,
defined as a distribution $P_F(\theta)$ over the behaviors of $F$. 

\subsection{Specification Learning}

Our Bayesian statistical framework builds a generative model of the form
$P(\theta, X, \Psi) = P(\theta | \Psi) P(X | \Psi) P(\Psi)$. This model
captures the intuition that every program is implementing some unknown
specification in the space of all specifications ($P(\Psi)$), which
determines the program's behavior ($P(\theta | \Psi)$) and
features ($P(X | \Psi)$).

Building this model requires data, in the form of a large corpus of
example programs. As in all statistical learning methods, we first
develop an appropriate statistical model, which is typically a
distribution family, and then \emph{learn} that model---choose the
parameters for the model family so they match reality---by training it on
data. To this end, $P(\theta, X, \Psi)$ also takes as
input a set of model parameters $\textbf{M}$.  Fully parameterized,
this distribution becomes:
\begin{equation}
P(\theta,  X, \Psi ; \textbf{M}) = 
P(\theta | \Psi; \textbf{M}) P(X | \Psi; \textbf{M}) P(\Psi ; \textbf{M})
\label{eqn:model}
\end{equation}

\ignore{
The distributions $P (X | \Psi; \textbf{M}) P(\Psi ; \textbf{M})$ and
$P(\theta | \Psi; \textbf{M})$ are learned by the underlying statistical
models that are used to instantiate our Bayesian framework.  In this
paper, we implement a Latent Dirichlet Allocation (LDA) model
(Section~\ref{subsec:lda}) and a topic-conditioned recurrent neural
network (Section~\ref{subsec:rnn}) respectively to learn these two
distributions.
}

The available data are then used to choose an appropriate set of
parameters \textbf{M}, using an optimization method such as maximum
likelihood. Suppose that we are given a large corpus of programs
$\{F_1,\ldots,F_N\}$, and for each program $F_i$ we have extracted
the pair $(X_{F_i}, \langle \theta_{i,1} ~ \theta_{i,2}, \ldots
\rangle)$ consisting of its feature set and a number of examples of its
behavior sampled from its behavior model.
Given this data, we would choose \textbf{M} that {\em maximizes} the
function:
\newcommand{\argmax}{\operatornamewithlimits{argmax}}
\begin{align}
\nonumber \prod_{i=1}^N \left(
\int_{\Psi} \left( \prod_{\theta_{i,j}} P(\theta_{i,j} |
\Psi; \textbf{M}) \right) P (X_{F_i} | \Psi; \textbf{M})
P(\Psi ; \textbf{M}) ~~ d\Psi \right) \end{align}

\noindent Note that we integrate out $\Psi$, since this is an
unseen random variable, as we typically do not know the value of the
precise specification associated with each code in the corpus.
\ignore{Solving
such a hidden-variable optimization problem may require use of an
appropriate algorithm, such as EM \cite{dempster1977maximum}. }
Once \textbf{M} is learned, the distribution would represent our prior
belief as to what the ``typical'' specification, behavior and features
look like, informed by the programs in the corpus.

\subsection{Anomaly Detection}

Suppose that we are given a new program $F$ and would like to obtain a
quantitative measure of the ``bugginess'' of $F$. 
%
On the one hand, since we already have learned a joint
distribution over behaviors, features and specifications,
$P(\theta, X, \Psi; \textbf{M})$, we can {\em condition} this distribution with
the newly observed $X_F$, to obtain the posterior:
$$P(\theta, \Psi | X_F ; \textbf{M}) = \frac{P(\theta, \Psi, X_F ;
\textbf{M})}{P(X_F ; \textbf{M})}$$
From Equation~\ref{eqn:model}, we have
$$P(\theta, \Psi | X_F ; \textbf{M}) = \frac{
P(\theta | \Psi; \textbf{M}) P(X_F | \Psi; \textbf{M}) P(\Psi ; \textbf{M})
}{P(X_F ; \textbf{M})}$$
Applying Bayes' rule to the term $P(X_F | \Psi ; \textbf{M})$ we get
$P(\theta, \Psi | X_F ; \textbf{M})$
\begin{align}
&= \frac{
    P(\theta | \Psi; \textbf{M}) \displaystyle{\frac{P(\Psi | X_F; \textbf{M}) P(X_F ;
    \textbf{M})}{P(\Psi ; \textbf{M})}} P(\Psi ; \textbf{M})
}{P(X_F ; \textbf{M})} \nonumber \\ \nonumber
&= P(\theta | \Psi; \textbf{M}) P(\Psi | X_F; \textbf{M})
\end{align}
From this, since we do not know the precise specification that $F$ is
implementing, we can integrate out $\Psi$ to obtain the (marginalized)
posterior distribution over behaviors:
\begin{equation}
    P(\theta | X_F; \textbf{M}) = \int_{\Psi} P(\theta | \Psi ;
    \textbf{M}) P(\Psi | X_F ; \textbf{M}) ~ d\Psi
\label{eqn:integral}
\end{equation}

\noindent
This particular form is very amenable to Monte Carlo integration, which
estimates an integral through random sampling.
Intuitively, it gives us a distribution over the program behaviors
$\theta$, that would be {\em anticipated}, given learned model parameters {\bf
M}, for a program with feature set $X_F$.

On the other hand, we have a distribution $P_F(\theta)$
over the {\em
actual} behaviors of $F$ when it is executed.
The final step is to then compare this actual distribution with the
anticipated distribution over behaviors, that is, $P_F(\theta)$ and
$P(\theta | X_F ; \textbf{M})$.
A measure such as the Kullback-Leibler (KL) divergence~\cite{KLD} between
distributions is appropriate here.  The KL-divergence between two
distributions $P_1$ and $P_2$ over the domain $i$ is a quantitative
measure defined as:
\begin{equation}
\kld{P}{Q} = \sum_i P_1(i) \log \frac{P_1(i)}{P_2(i)}
\label{eqn:kld}
\end{equation}

Using this measure, we can compute the {\em anomaly score} of $F$ by setting
$P_1$ and $P_2$ to the distributions $P_F(\theta)$ and $P(\theta | X_F;
\textbf{M})$ respectively, and ranging $i$ over the domain of all possible
program behaviors in the language $\Sigma^*$:
\begin{equation}
    \sum_{\theta \in \Sigma^*} P_F(\theta) \log
\displaystyle{\frac{P_F(\theta)}{P(\theta | X_F; \textbf{M})}}
\label{eqn:score}
\end{equation}

\paragraph{Choosing an Abstraction}

When instantiating the framework, the exact form of the feature set
$X_F$ must be chosen with some care.  If the feature set $X_F$ does
not provide any abstraction for the program (in the extreme case,
$X_F$ is merely the program itself) and the model and learner are
arbitrarily powerful, then $P(\theta|X_F;\textbf{M})$ (Equation 2)
could, in theory, describe the compiler and symbolic executor
used to produce the training data. This would mean that the KL divergence
(Equation~\ref{eqn:kld}) is zero for any program.

When applying the framework to a problem, we protect against this
possibility by choosing a feature set $X_F$ that abstracts the program
to an appropriate level for the debugging task. For example, when
debugging API usage,
it makes sense to choose $X_F$ as the bag of API calls made by the
code. This ensures that $P(\theta|X_F;\textbf{M})$ is limited to
attaching probabilities to various sequences that can be made out of
those calls, and it is impossible for the learner to ``learn'' to
compile and execute a program.

\section{Instantiation of the Framework}\seclabel{instantiation}

In this section, we present an instantiation
of our framework, and discuss practical implementation challenges.

\subsection{Probabilistic Behavior Model $P_F(\theta)$}
\label{subsec:semantics}

First, our instantiation includes 
a definition of the probabilistic behavior model $P_F(\theta)$. This
definition relies on the abstraction of programs as {\em generative
probabilistic automata}~\cite{GPA1,GPA2}.

\paragraph{Program Model.} A generative probabilistic automaton
is a tuple $F = \langle \states, \Sigma, q_0, Q_A, \trans \rangle$ where
$\states$ is a finite set of states, $\Sigma$ is the alphabet of
observable symbols that was introduced earlier, $q_0 \in \states$ is
the {\em initial state}, $Q_A \subseteq \states$ is a set of {\em
  final or accepting states}, and
$\trans: \states \times \Sigma \times \mathbb{R}_{(0,1]} \times
\states$
is a {\em transition relation}. We have $\trans(q_i, s, p, q_j)$ if
the automaton can transition between states $q_i$ and
$q_j$ with a probability $p \in (0,1]$, generating the symbol $s$. (We 
write 
$\transition{q_i}{q_j}{s}{p}$ if such a transition exists.)  
Transitions with probability 0, or
infeasible transitions, are excluded from the automaton.

A program in a high-level language is transformed into the above
representation through {\em symbolic execution}~\cite{King76SE},
during a preprocessing phase. Symbolic execution runs a program with symbolic inputs and
keeps track of {\em symbolic states}, which are analogous to a
program's memory. The symbolic states encountered become the states
$\states$, and the accepting states $Q_A$ are typically the states at
a final location (or some location of interest) in the
program. Unbounded loops can be handled by imposing a bound on
symbolic loop unrolls, or through a predicate abstraction of the
program to make variable domains finite. The detection of infeasible
states---in general an undecidable problem---depends on the underlying
theorem prover used by symbolic execution.

As symbolic execution is a standard method in formal
methods~\cite{tracercav12,KLEE,JPFSE}, this section only
gives an example of the method's use. \ignore{More details appear in the
supplementary material (Appendix~\ref{sec:symexec}).} As it is applied at a
preprocessing level, we often use the
term ``program'' to refer to an automaton generated via symbolic
execution, rather than a higher-level program to which preprocessing
is applied.

\paragraph{Semantics.}
A {\em run} $\pi$ of $F$ is defined as a finite sequence of transitions
$\transition{q_0}{q_1}{s_1}{p_1} \transition{}{}{s_2}{p_2} \cdots
\transition{}{q_n}{s_n}{p_n}$
beginning at the initial state $q_0$.  $\pi$ is {\em accepting} if
$q_n \in Q_A$.  The probability of a run
$P(\pi) = \prod_{i=1}^{n} p_i$. Every run $\pi$ generates a 
behavior $\theta \in \Sigma^*$, denoted as
$[|\pi|] = s_1 s_2 \cdots s_n$.  Let $\Pi_F$ be the set of all
accepting runs of $F$, and $\Pi_F(\theta) \subseteq \Pi_F$ be the set
of all accepting runs $\pi$ such that $[|\pi|] = \theta$.
The {\em probabilistic behavior model} 
$P_F(\theta): \Sigma^* \rightarrow [0,1]$ is:
\begin{equation}
P_F(\theta) = \displaystyle{\frac{1}{Z}} \sum_{\pi \in \Pi_F(\theta)} P(\pi).
\label{eqn:semantics}
\end{equation}
Here, $Z = \sum_{\pi \in \Pi_F} P(\pi)$
 is a {\em normalization factor}.
%

It is easy to see that $P_F(\theta)$ defines a probability
distribution over behaviors.
To generate a ``random'' behavior of $F$, we simply sample from the
distribution $P_F(\theta)$. 


\paragraph{Features.} Given a program $F$, the feature set $X_F$ is
defined as $\{s ~|~ \transition{q_i}{q_j}{s}{p} \in \trans\}
\setminus \{\epsilon\}$,
i.e., the set of all non-empty symbols in the transition system of $F$.

\begin{figure}
 \vspace{-0.05in}
    \hspace{0.2cm}\includegraphics[scale=0.59]{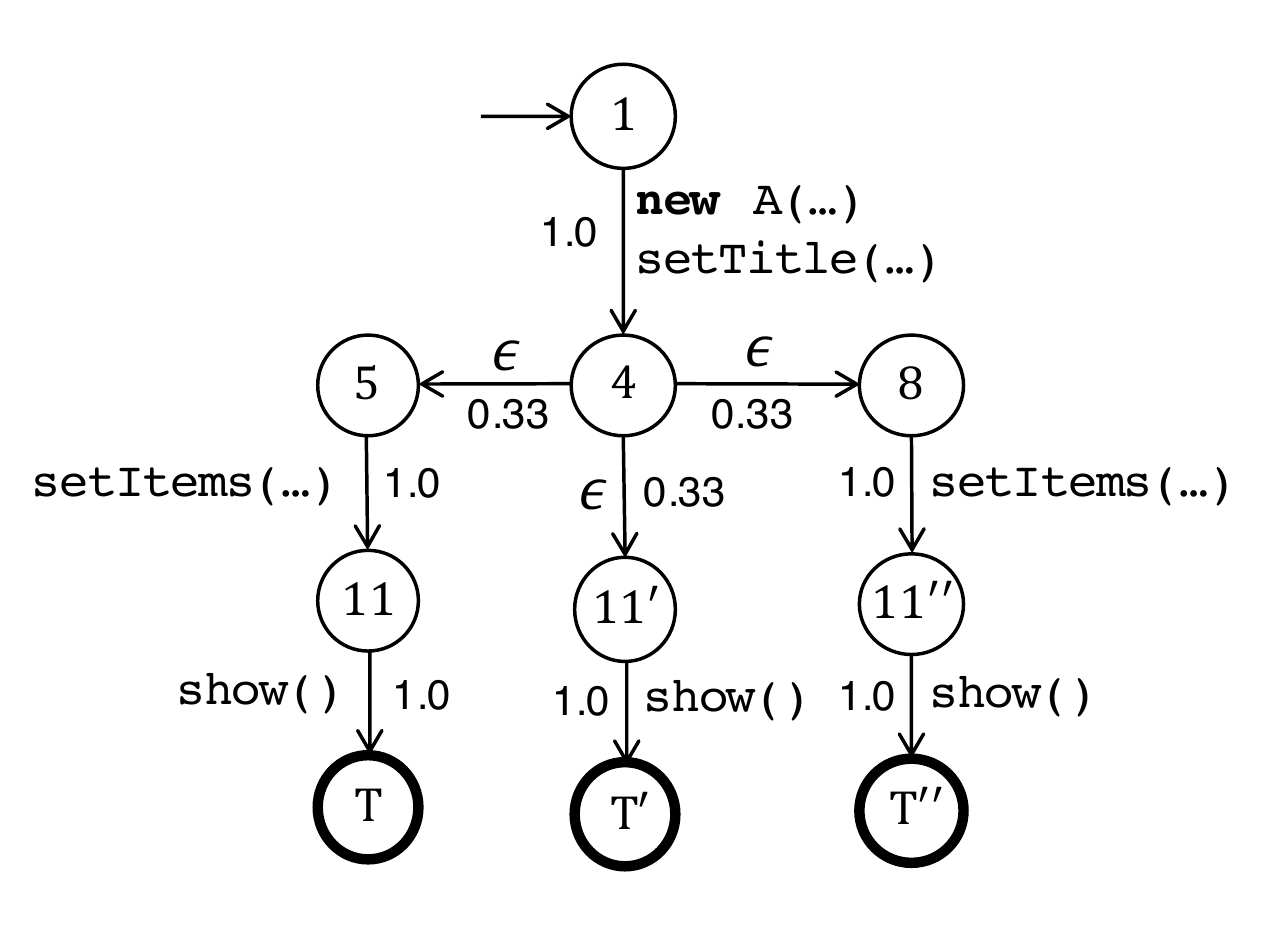}
\vspace{-0.05in}
    \caption{Automaton for the example in
    Figure~\ref{fig:ui-bug}(b)(i)}
    \label{fig:automaton}
\vspace{-0.1in}
\end{figure}

\paragraph{Example.}
The automaton model for the code in Figure~\ref{fig:ui-bug}(b)(i)
is shown in Figure~\ref{fig:automaton}. Each ``state'' in
the automaton is labeled with a program location, with multiple
instances of the same location being primed. The initial state is the
first location, and the accepting states, in bold, are all instances of
a (special) terminal location $T$ in the program. The transitions follow the
structure of the code (for brevity, we collapse sequential statements into a
single transition), emitting as symbols API methods called at each
location.

Note that we gave a uniform probability at each state to transition to
the next possible states, but this can be controlled through other
means. For instance, one can apply {\em model counting} on a branch
condition and compute the probability of the program executing one
branch over another. Such a definition is not necessarily a better
choice than ours, as it would assign low
probabilities to corner cases that get triggered on a small number of
inputs but are often of interest to users of static
analysis. The two definitions simply make different tradeoffs. We go
with a uniform distribution at branches 
because it is simpler and worked well in
our experiments. 


$\{{\tt new ~ A()}, {\tt setTitle(\ldots)},$ ${\tt setItems(\ldots)}, {\tt
show()}\}$ is the feature set for this program.  There are three accepting
runs of $F$, and two behaviors generated by these accepting runs:

\vspace{1mm}
\begin{tabular}{l}
$\theta_1 = {\tt new~A()} ~~ {\tt setTitle(\ldots)} ~~ {\tt setItems(\ldots)} ~~ {\tt show()}$\\
$\theta_2 = {\tt new~A()} ~~ {\tt setTitle(\ldots)} ~~ {\tt show()}$
\end{tabular}

\vspace{1mm}\noindent
We have $Z = 1.0$, the sum of the probabilities of all accepting runs.
Hence, $P_F(\theta_1) = (0.33 + 0.33)/1.0 = 0.66$ and
$P_F(\theta_2) = 0.33$. 

Assume now that after training on a large number of behaviors, the
statistical model had learned that
conditioned on specifications such as $\langle 0.98, 0.01, 0.01 \rangle$
(that gave a high probability to the first topic), program
behaviors tend to always add a title and items to dialog boxes 
before being shown.  This might result in the behavior $\theta_1$ having
a very high probability, say 0.99, and all other behaviors having a very
low probability.
Particularly, a behavior that only calls ${\tt setTitle}$ without ${\tt
setItems}$ would be assigned a very low probability, say, $10^{-5}$.
In our program $F$, we saw that $P_F(\theta_1) = 0.66$ and
$P_F(\theta_2) = 0.33$, and the probability of any other $\theta$ is
0. Thus, the anomaly score of $F$ is:
$0.66 ~ \log \frac{0.66}{0.99} + 0.33 ~ \log \frac{0.33}{10^{-5}} =
3.16$
Suppose now, that the state $11'$ in the program
model was infeasible. Then, both accepting runs in the model would only
generate $\theta_1$, and so $P_F(\theta_1) = 1$. The anomaly score of
this ``correct'' program would then be $\log \frac{1}{0.99} = 0.01$.

\subsection{Topic Models for $P(\Psi , X; \textbf{M})$}
\label{subsec:lda}
Topic models are used in natural language processing to automatically
extract topics from a large number of ``documents'' containing textual
data as words. In our case, documents are the feature sets
$X$, words are symbols from the alphabet (vocabulary) $\Sigma$, and the
topic distribution of a document is its unknown specification $\Psi$.

\ignore{
\begin{figure}
\hspace{1cm}
\includegraphics[scale=0.5]{LDA.pdf}
\caption{Variables in LDA, where $w$ is the only observable.}
\label{fig:LDA}
\end{figure}
}

LDA~\cite{LDA} is a popular topic model that models the generative
process of documents in a corpus where each document $X_{F_i}$ contains
a bag of words. The inputs to LDA are the number of topics to be
extracted $K$, and two hyper-parameters $\alpha$ and $\eta$.
LDA models a document as a distribution over topics, and a topic as a
distribution over words in the vocabulary. An LDA model is 
characterized by the variables: (i) $\alpha$ and $\eta$,
hyper-parameters of a Dirichlet prior that chooses the topic
distribution of each document and the word distribution of each topic,
respectively (ii) $\Psi_{F_i}$, the topic distribution of document
$X_{F_i}$, (iii) $\beta_k$, the word distribution of topic $k$.

The result of training an LDA model is a learned value for all the latent variables
$\alpha$, $\eta$, $\Psi_{F_i}$ and $\beta_k$, which forms our model
parameter \textbf{M}.
During inference, we are given a document $X_F$, and we would like to
compute the posterior distribution $P(\Psi | X_F; \textbf{M})$. Since
LDA has already learned a joint distribution $P(\Psi, X ; \textbf{M})$,
this is simply a matter of conditioning this distribution with the newly
observed $X_F$ to get a posterior distribution over $\Psi$, which is
often approximated through a technique called {\em Gibbs
sampling}~\cite{gibbsSampling}. \ignore{See the supplementary material (Appendix~\ref{sec:gibbs})
for more details.}

\subsection{Recurrent Neural Networks for $P(\theta | \Psi; \textbf{M})$}
\label{subsec:rnn}

\ignore{
\begin{figure}
\hspace{0.5cm}
\includegraphics[scale=0.47]{RNN.pdf}
\caption{Topic-conditioned Recurrent Neural Network}
\label{fig:RNN}
\vspace{-0.1in}
\end{figure}
}

\ignore{
The final task is to define the distribution over behaviors, given a
$\Psi_F$ sampled from the LDA model. For this purpose, we employ a
topic-conditioned RNN~\cite{TCRNN} implemented as a character-level
model.}
\ignore{We refer to the work~\cite{vechevPLDI14} that
elucidates these models well, nevertheless we provide a description here
for completeness.}

Neural networks have been used to solve classification problems such as
image recognition and part-of-speech tagging.  These problems involve
classifying an input {\bf x} into a set of (output) classes {\bf y},
using the conditional distribution $P(\textbf{y} | \textbf{x},
\textbf{M})$ where {\bf M} is the set of neural network parameters.
\ignore{For instance, if {\bf x} is a handwritten image and {\bf y} is
a vector of numbers $\langle 0,\ldots,9 \rangle$, then this distribution
gives the probability of {\bf x} being each of the numbers in {\bf y}.
Character-level models turn such a classification task into a {\em
generative} task -- one that generates outputs -- by some clever
encoding of the input {\bf x} and classes {\bf y}.}

Suppose that {\bf x} is a given sequence of symbols (characters)
$s_1 s_2 \ldots s_{t-1}$ where each symbol is from the alphabet $\Sigma$,
and we would like the model to generate the next symbol $s_t$. We can
cast this generative problem as a classification task by creating $x_1
x_2 \ldots x_{t-1}$ where each $x_k$ is the {\em one-hot vector} of $s_k$
\ignore{-- that is, a vector where exactly one element corresponding
to the index of $s_k$ in $\Sigma$ is one, and the rest are zeroes -- }and
querying the model to ``classify'' the sequence $x_1 x_2 \ldots x_{t-1}$
into $|\Sigma|$ classes. The output vector $y_t$ is then interpreted as
a distribution over $\Sigma$, from which a symbol $s_t$ can be
sampled~\cite{bengio03neural}.
Let us denote the probability of a symbol $s$ given by the output
distribution $y_t$ as $y_t(s)$.

A topic-conditioned neural network~\cite{topicRNN} takes, in addition to {\bf x}, an
input $\Psi$ representing the topic distribution of a
document obtained from a topic model. \ignore{Note that $\Psi$ cannot be encoded as a
one-hot vector (it is a vector of reals), and is {\em given} -- it is
not part of the output of the neural network.}
To handle unbounded length input sequences, a {\em recurrent} neural
network is used. An RNN uses a hidden state
to neurally encode the sequence it has seen so far.
At time point $t$, the hidden state $h_t$ and the output $y_t$ are computed
as:
\begin{equation}
h_t = f(\textbf{W} h_{t-1} + \textbf{V} \Psi + \textbf{U} x_t + \textbf{b}_h), ~~~
y_t = g(\textbf{T} h_t + \textbf{b}_y)
\label{eqn:y}
\end{equation}
where $\textbf{W}$, $\textbf{V}$, $\textbf{U}$ and  $\textbf{T}$ are the
weight matrices of the RNN, $\textbf{b}_h$ and $\textbf{b}_y$ are the bias 
vectors of the hidden states and outputs respectively, $f$ is a
non-linear {\em activation} function such as the sigmoid, and $g$ is a
{\em softmax} function that ensures that the output is a 
distribution.

\ignore{
$$f(z) = \frac{1 - e^{-2z}}{1 + e^{-2z}} ~~~~~~~~ g(z_j) =
\frac{e^{z_j}}{\sum_k e^{z_k}}$$}

Training the model involves defining an error function between the
output of the RNN and the observed output in the training data.
Specifically, if the training data is of the form $(X_{F_i}, \langle
\theta_{i,1}, \theta_{i,2}, \ldots \rangle)$, then each training step of
the RNN will consist of the input {\bf x} being $\theta_{i,j}$, target
output {\bf y} being $\theta_{i,j}$ shifted by one position to the left
(since at time point $t$ the output $y_t$ is interpreted as the
distribution over the {\em next} symbol in the sequence), and $\Psi$ being a sample
from $P(\Psi | X_F; \textbf{M})$ given by the trained topic model.
A standard error function such as cross-entropy between the output of
the RNN and the target output can be used.

Since the error function and all non-linear functions used in the RNN are
differentiable, training is done using stochastic gradient descent.  The
result of training is a learned value for all matrices in
the RNN, which together form a part of our model parameter {\bf M}.

During inference, we are given a $\Psi$ and a particular $\theta =
s_1, \ldots, s_n$, and would like to compute $P(\theta | \Psi;
\textbf{M})$. This is straightforward: we set $x_t$ as the
one-hot vector of $s_t$ for $1 \leq t \leq n$. Then,
$P(\theta | \Psi; \textbf{M}) = \prod_{t=1}^{n-1} y_t(s_{t+1})$
where $y_t$ is computed using Equation~\ref{eqn:y}.

\subsection{Estimation of the Anomaly Score}
There are two difficulties associated with computing the anomaly score
in our instantiation of the framework. The first is that in general, the
computation given in Equation~\ref{eqn:score} requires summing over a
possibly infinite number of program behaviors $\theta$, which is not
feasible. Second, it also requires computing $P(\theta|X_F;\textbf{M})$,
which in turn requires integrating out the unknown specification $\Psi$
(Equation~\ref{eqn:integral}).

Both of these difficulties can be addressed via sampling.
We note that in general, to estimate a summation of
the form $\sum_{i \in I} P_1(i) P_2(i)$ where $P_1(i)$ is a probability
mass function over the (possibly) infinite domain $I$ and $P_2$ is a
function on $I$, it suffices to take a number of samples $i_1,
i_2, \ldots, i_m \sim P_1(i)$.  One can then use: $$\sum_{i \in I} P_1(i)
P_2(i) \approx \sum_{k = 1}^m \frac{1}{m} P_2(i_k)$$ as an unbiased
estimate for the desired sum.  It is well known from standard sampling
theory that the variance of this estimator, denoted as $\sigma^2$,
reduces linearly as $m$ increases. 

We can apply this estimation process to estimate the anomaly score for
$F$ by letting the domain $I$ be the set of all possible behaviors in
$\Sigma^*$, and
sampling a large number of behaviors with probability proportional to
$P_F(\theta)$, then letting $P_2(\theta) = \log
(P_F(\theta)) - \log(P(\theta | X_F; \textbf{M}))$ and using the
estimator described above.  We can keep sampling until the variance of
the estimate is sufficiently small.  

Fortunately, sampling a behavior from the distribution $P_F(\theta)$ is
easy: we can use rejection sampling~\cite{vonNeumann51} to sample an accepting run $\pi$ of
$F$ and then simply obtain its behavior $\theta =
[|\pi|]$.
However we do not yet have a complete solution to our problem. The
difficulty is that for a sampled behavior $\theta$, it is not possible
to compute $P_2(\theta)$ easily because of two reasons.  First, the term $P_F(\theta)$
(Equation~\ref{eqn:semantics}) requires summing over possibly infinite
number of accepting runs $\Pi_F$, and second, there is the aforementioned problem
that computing $P(\theta | X_F; \textbf{M})$ requires integrating over the unseen
$\Psi$ value.

To handle this, we extend our sampling-based algorithm. Rather
than just sampling a set of behaviors, we sample the set $I$ of
$(\theta, \Pi_F', \psi_F)$ triples, where $\Pi_F'$ is itself a set of accepting runs
of $F$ sampled using the same method, and $\psi_F$ is a set of values for
$\Psi$ sampled from $P(\Psi|X_F; \textbf{M})$. The latter set of samples can easily
be obtained via Gibbs sampling.
One could then estimate the divergence as: 
\begin{align}\frac{1}{|I|} \sum_{\substack{(\theta, \Pi_F', \psi_F) \\ \in I}}
        \log \left(\sum_{\pi \in \Pi_F'(\theta)} \frac{1}{|\Pi_F'|}\right) - \nonumber
        \log \left( \sum_{\Psi \in \psi_F} P(\theta | \Psi; \textbf{M})\right) \nonumber
\end{align}
\noindent where $\Pi_F'(\theta)$ is the set of paths $\pi \in \Pi_F'$
such that $[|\pi|] = \theta$. The sum inside the first logarithm is 
estimating the fraction of sampled accepting runs whose behavior is
$\theta$, thereby estimating $P_F(\theta)$ through sampling, and the sum inside of the
second logarithm is estimating $P(\theta | X_F; \textbf{M})$.

The problem is that this estimate will be biased, since one cannot commute
the expectation operator with a logarithm.
That is:
\begin{align}
    E \left[ \log (\sum_{\pi \in \Pi_F'(\theta)} \frac{1}{|\Pi_F'|}) 
\right] \neq 
\log (E \left[ \sum_{\pi \in \Pi_F'(\theta)} \frac{1}{|\Pi_F'|} \right] )\nonumber 
\end{align}

\noindent where $E[Y]$ for a random variable $Y$ denotes the expectation
of $Y$. A similar problem exists for the second summation used to estimate
the logarithm of $P(\theta | X_F; \textbf{M})$.
Intuitively, this bias is not surprising, since an over-estimate for the
probability $P_F(\theta)$ by some constant amount is likely to
have little effect on an estimate of the logarithm of the probability.
However, an under-estimate by the same amount can cause a radical
reduction in the estimate of the logarithm, and we expect a negative
bias.

\ignore{we refer the reader to the supplementary material (Appendix~\ref{sec:bias}) for
details} 

A sampling-based estimate for this bias can be computed
using a Taylor series expansion about the expected value of
the biased estimator, which obtains an expression for the bias
in terms of the central moments of a Normal distribution; estimating
those moments leads to an estimate for the bias. Assume that
this estimator is encapsulated in a procedure
$bias(\theta, \Pi_F',\psi_F)$ that computes the bias
of an estimate.
Our final estimate for the anomaly score is:
\begin{align}\frac{1}{|I|} \sum_{(\theta, \Pi_F', \psi_F) \in I}
        \log \left(\sum_{\pi \in \Pi_F'(\theta)} \frac{1}{|\Pi_F'|}\right)
        - \nonumber
	\log \left( \sum_{\Psi \in \psi_F} P(\theta | \Psi; \textbf{M})\right)
    \\
        - bias(\theta, \Pi_F',\psi_F) \nonumber
\end{align}

\section{Evaluation}\seclabel{eval}

\lstset{mathescape=true,numbers=none,basicstyle=\fontsize{7}{7}\selectfont\ttfamily}
\begin{figure}
    \begin{tabular}{lll}
        {\sf Topic 1} & {\sf Topic 2} \\
        \hspace{-1cm}
        \begin{lstlisting}
        A.setMessage(int)
        A.setTitle(int)
        new A(Context)
        \end{lstlisting}
        &
        \hspace{-1cm}
        \begin{lstlisting}
        A.setPositiveButton(String,$\ldots$)
        A.setNegativeButton(String,$\ldots$)
        A.setMessage(String)
        \end{lstlisting}
        \\ \\
        {\sf Topic 3} & {\sf Topic 4} \\
        \hspace{-1cm}
        \begin{lstlisting}
        A.setView(View)
        new A(Context)
        A.setTitle(String)
        \end{lstlisting}
        &
        \hspace{-1cm}
        \begin{lstlisting}
        A.setItems(String[],$\ldots$)
        A.setNeutralButton(int,$\ldots$)
        A.show()
        \end{lstlisting}
        \\ \\
        {\sf Topic 5} & {\sf Topic 6} \\
        \hspace{-1cm}
        \begin{lstlisting}
        C.getInstance(String)
        C.init(int,Key,$\ldots$)
        C.doFinal(byte[])
        \end{lstlisting}
        &
        \hspace{-1cm}
        \begin{lstlisting}
        B.connect()
        B.getInputStream()
        B.getOutputStream()
        \end{lstlisting}
    \end{tabular}
    \caption{Top-3 methods from topics extracted by LDA
        (${\tt A}$ = ${\tt AlertDialog.Builder}$, ${\tt B}$ = ${\tt
    BluetoothSocket}$,  ${\tt C}$ = ${\tt Cipher}$)}
    \label{fig:topics}
\end{figure}

In this section, we present results of experimental evaluation of our
method on the problem of learning specifications and detecting anomalous
usage of Android API in Android apps. Specifically, we seek to answer
the following questions:
\begin{itemize}
    \item[(1)] Can we find useful {\em de facto} specifications
        followed by Android developers (Section~\ref{subsec:learning})?
    \item[(2)] Using the specifications, can we find possible bugs in
        the usage of the Android API in a corpus
        (Section~\ref{subsec:anomaly})?
    \item[(3)] How does specification learning help in anomaly
        detection (Section~\ref{subsec:knn})?
    \item[(4)] How does the Bayesian framework help in handling
        heterogeneity in the specifications (Section~\ref{subsec:mining})?
\end{itemize}

\ignore{This application domain is a
good fit for our method because there exist a number of usage practices
that are unspecified by the API but are {\em de facto} followed by
Android developers. Our framework would be able to automatically learn
these unknown specifications and can be applied to detect deviant
behaviors in new Android apps. The availability of large corpora of
these apps also helps in the training phase.}

\subsection{Implementation and Experimental Setup}
\label{subsec:implementation}

We first set up the practical
environment for the experiments. We implemented our method in a tool
named Specification Learning Tool, or \salento.  \salento\ uses {\sf
soot}~\cite{soot} to implement symbolic execution and
transform code in an Android app into our automaton model for programs, {\sf
TensorFlow}~\cite{tensorflow} to implement the topic-conditioned RNN,
and {\sf scipy}~\cite{scipy} to implement LDA.  \salento\ builds a
coarse model of the Android app life-cycle by collecting all entry points
in the application which are callback methods from the Android kernel.
\ignore{To increase the precision of the interprocedural symbolic execution,}
\salento\ also uses {\sf soot}'s Class Hierarchy Analysis and Throw
Analysis and obtains an over-approximation of the set of possible
call or exception targets, and {\sf soot}'s built-in constant propagator
to detect infeasible paths.

In addition to API methods in $\Sigma$, \salento\ also collects some
semantic information about the state of the program when an API call is
made. This is done through the use of simple Boolean {\em
predicates} that capture, for example, constraints on the arguments of
a call, or record whether an exception was thrown by the call. This
allows us to learn specification on more complex programming constructs.

The training corpus consisted of 500 Android apps
from~\cite{androiddrawer}, and the testing corpus consisted of 250
apps from~\cite{fdroid}. The two repositories did not overlap, perhaps
since the latter is open-source and the former is not.  We conducted
experiments on three APIs used by Android apps: builders for alert
dialog boxes (${\tt android.app.AlertDialog.Builder}$), bluetooth sockets
(${\tt android.bluetooth.BluetoothSocket}$) and cryptographic ciphers
(${\tt javax.crypto.Cipher}$). From the two repositories, we created
about 6000 and 1800 automata models (henceforth just called programs)
respectively. While doing so, we set the accepting location of the program as
various locations of interest, that is, locations where a method in
one of these APIs was invoked. This helps in {\em localizing} an
anomaly to a particular location.
All experiments were run on a 24-core 2.2 GHz machine with 64 GB
of memory and an Nvidia Quadro M2000 GPU. 

\paragraph{Remark.} The three APIs were chosen to represent common yet
varied facets of a typical Android app (UI, functionality, security).
Evaluating on more APIs is not a fundamental limitation of our method or
implementation. Rather the limiting factor is, as we will see soon, the manual
effort that has to be spent in order to triage more anomalies and report
precision/recall.

\subsection{Specification Learning}
\label{subsec:learning}

With a goal to discover specifications of Android API usage,
we applied LDA on the training corpus of programs, where the
alphabet $\Sigma$ consisted of 25 methods from the three APIs. We used
$\alpha = 0.1$ for each topic, and $\eta = 1/|\Sigma|$
for all words in a topic. \ignore{The feature set $X_F$
consisted of the set of API methods called in the program $F$.}
Running LDA with $15$ topics ($K$) took a few seconds to
complete. Figure~\ref{fig:topics} shows the top-3 words (methods) from
six topics extracted from the corpus that we picked to exemplify. At a
first glance, it may seem that LDA is simply categorizing
methods from different APIs into separate topics, which can raise the
question of why we need topic models if we already
knew the APIs beforehand.

LDA, however, does more than that. {\sf Topic 1} and {\sf Topic
2} contain methods from the same API but, interestingly, different
polymorphic versions with ${\tt int}$ and ${\tt String}$
arguments. The model has discovered that the
polymorphic versions fall under separate topics, meaning that they are
{\em not often used together in practice}. Indeed, some Android apps
declare all resources they need in a separate XML file, and provide the
resource ID as the ${\tt int}$ argument. Other apps do not make
use of this feature and instead directly provide the string to use in
the dialog box. Therefore, it makes sense that
an app would seldom use both versions together.
Similarly, {\sf Topic 3} also contains methods from the same API,
however it describes yet another way to create dialog boxes. Note the
lack of the ${\tt setMessage}$ method in this topic, as the message
would already have been enveloped in the ${\tt View}$ passed to ${\tt
setView}$ (using both methods together can lead to the display of
corrupted dialog boxes as shown in Figure~\ref{fig:ui-bug}(a)(ii)).

As these examples show, the topic model can expose specifications of how
methods in an API, or even different APIs, are used together in
practice.

\subsection{Anomaly and Bug Detection}
\label{subsec:anomaly}

\begin{figure*}
\begin{tabular}{cccc}
    \begin{minipage}{0.25\textwidth}
    \hspace{-1cm}\includegraphics[scale=0.3]{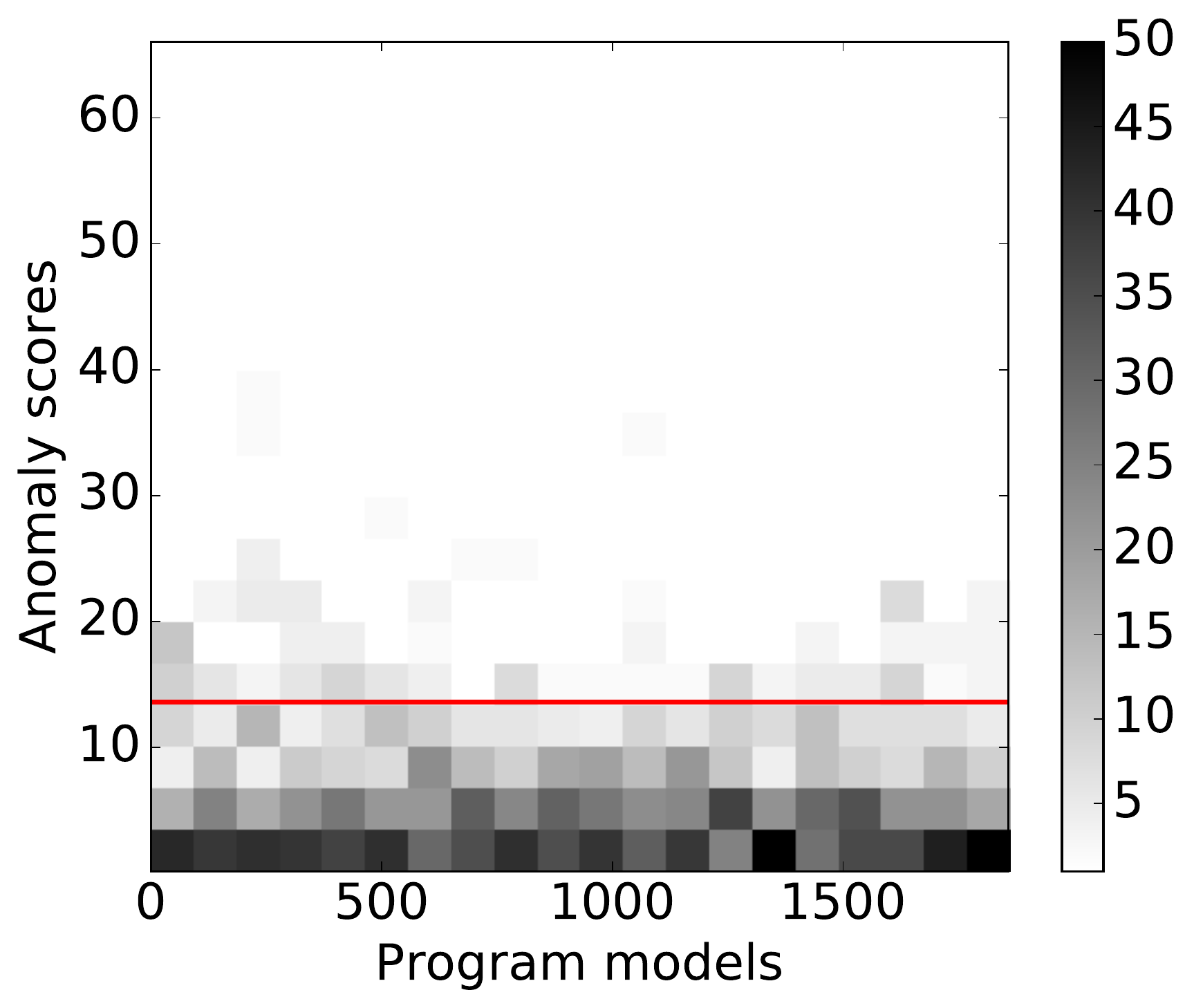}
    \end{minipage}
    &
    \begin{minipage}{0.2\textwidth}
    \includegraphics[scale=0.4]{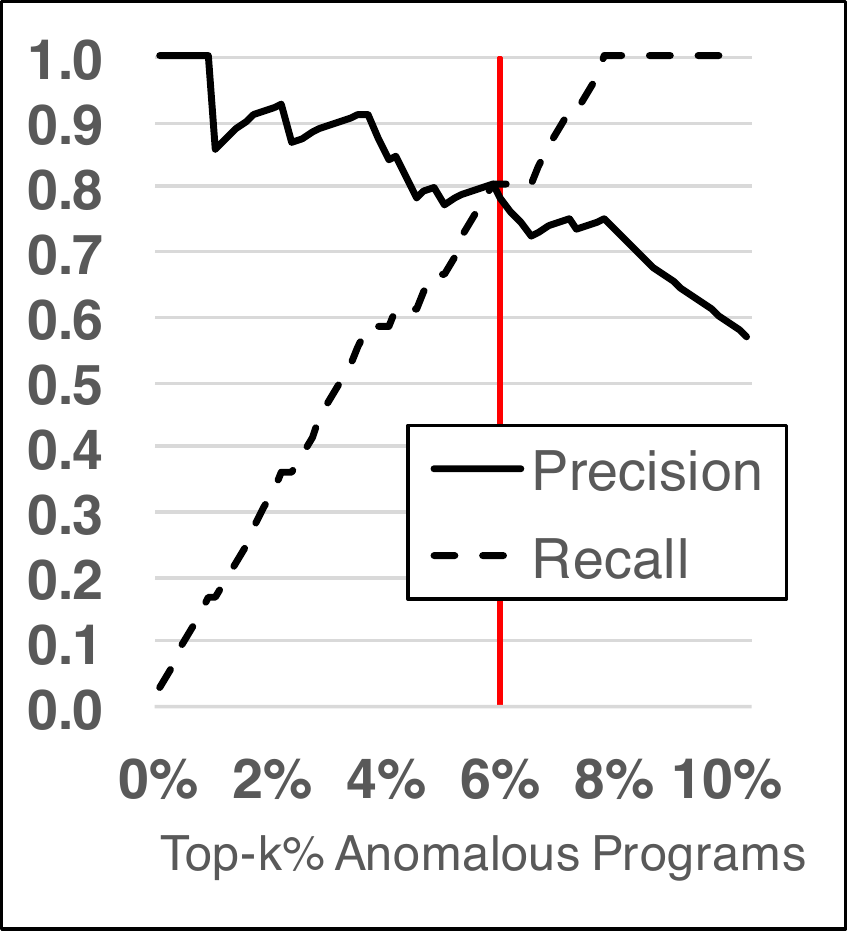} 
    \end{minipage}
    &
    \begin{minipage}{0.17\textwidth}
    \hspace{-0.3cm}\includegraphics[scale=0.38]{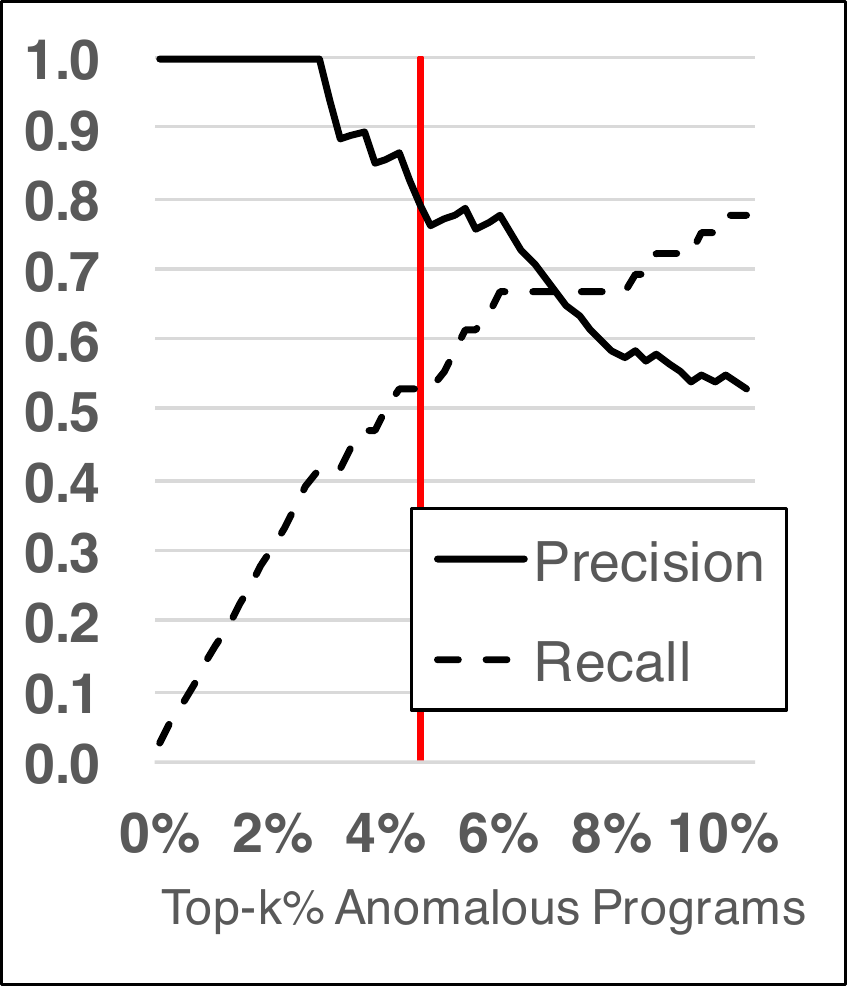}
    \end{minipage}
    &
    \begin{minipage}{0.25\textwidth}
    \hspace{0.2cm}\includegraphics[scale=0.4]{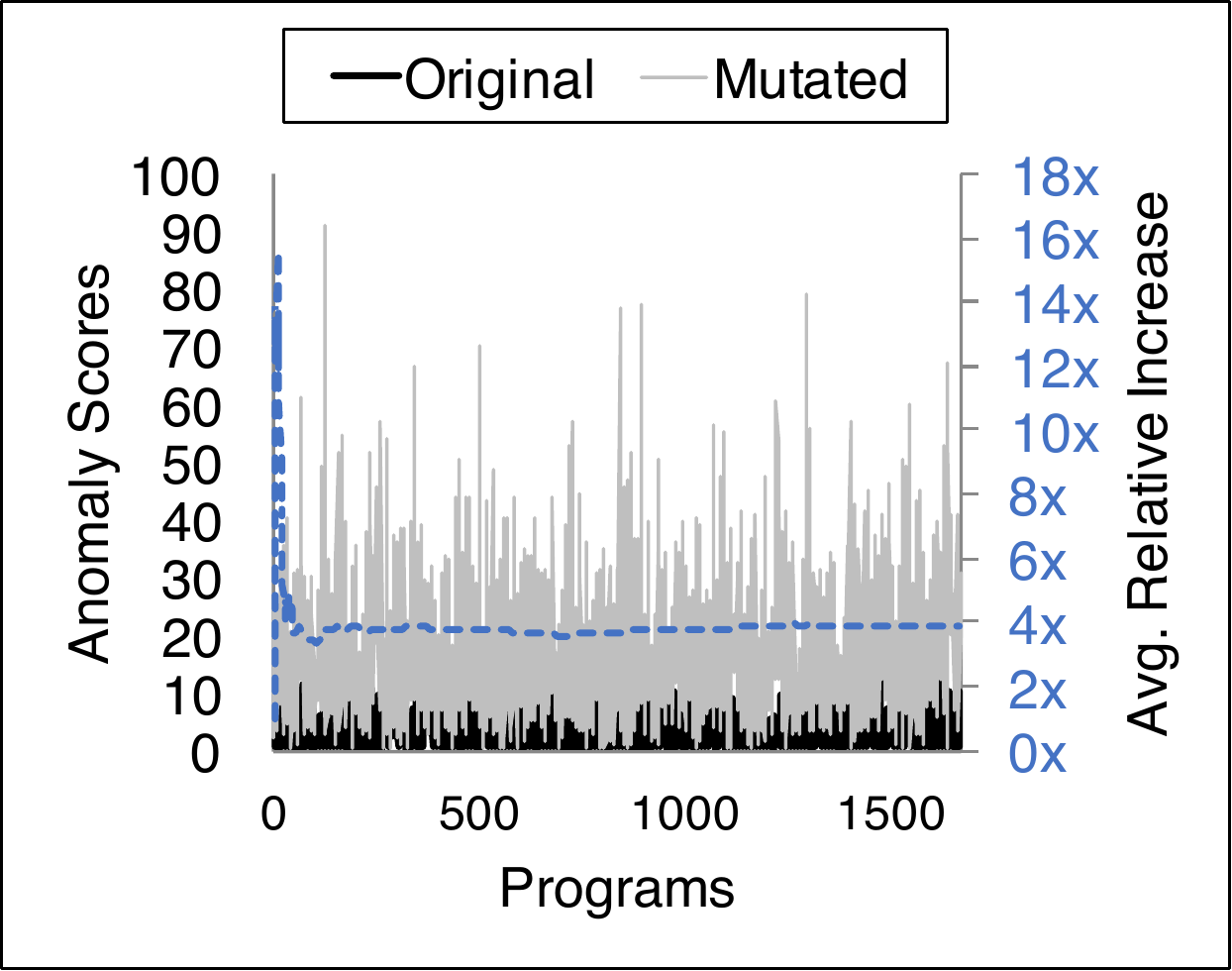}
    \end{minipage}
    \\ (a) & (b)(i) & (b)(ii) & (c)
\end{tabular}
\caption{(a) Histogram of anomaly score values, (b) Precision-recall for the
    possible bugs in Figure~\ref{fig:anomalies} for (i) Bayesian model (ii)
    non-Bayesian model, and (c) Anomaly scores of remaining 90\% programs before and
    after mutation}
\label{fig:results}
\end{figure*}

\begin{figure}
    \hspace{-0.3cm}
    \begin{tabular}{|r|r|r|p{0.1mm}p{5cm}|} 
        \hline
        {\sf \#} & {\sf Count} & {\sf Avg} & & \multicolumn{1}{c|}{\sf Anomaly} \\ 
                 &             & {\sf Score} && \\ \hline
        1 & 2 & 43.7 & ${\tt C}$ & Single crypto object used to encrypt/decrypt multiple data \\
        2 & 1 & 37.5 & ${\tt B}$ & Connecting to the same socket more than once \\
        3 & 1 & 24.7 & ${\tt B}$ & Attempt to close unopened socket \\
        4 & 16 & 22.1 & ${\tt A}$ & Using ${\tt String}$ and ${\tt int}$ polymorphic methods together \\
        5 & 6 & 21.8 & ${\tt C}$ & Crypto object created without specifying mode \\
        6 & 6 & 21.6 & ${\tt A}$ & Using ${\tt setMessage}$ with ${\tt setView}$ \\
        7 & 1 & 19.8 & ${\tt A}$ & Dialog displayed without message \\
        8 & 1 & 19.3 & ${\tt B}$ & Failed socket connection left unclosed \\
        9 & 1 & 16.5 & ${\tt A}$ & Unusual button text \\
       10 & 1 & 15.7 & ${\tt A}$ & Dialog displayed without buttons \\
        \hline
    \end{tabular}
    \caption{Anomalies that are possible bugs, found in the top 10\% of anomalous programs}
    \label{fig:anomalies}
\end{figure}

To evaluate our method on anomaly detection, we first
trained the topic-conditioned RNN on 60,000 behaviors sampled from the training
programs. Training took 20 minutes to complete.
We then computed anomaly scores for the 1800 programs in our
testing corpus. The time to compute each score was around 2-3 seconds.

The histogram of scores, in Figure~\ref{fig:results}(a), shows a high
concentration of small values, such as 5 or less, and a very low
concentration of high values. We chose to further investigate programs
appearing in the top 10\% of anomaly scores (above the red
line) for possible bugs. Specifically, since each program 
provides a localization to a location in the app (through its
accepting states), we investigated the behaviors that were sampled
from the program's probabilistic behavior model, that would have determined its
anomaly score.

Our definition of a ``possible bug'' is based on the following:
is a behavior an instance of Android API usage that is questionable
enough that we would expect it to be raised as an issue in a formal code
review? Note that an issue raised in a code review may relate to a
design choice and not necessarily cause the program to crash (an unusual
button text, for example). Nonetheless, such an issue would
be raised and likely fixed by engineers examining a code.

One problem with counting an anomaly as a possible bug is that multiple
anomalies in an app can have the same ``cause''---an incorrect statement
or set of statements in the code---and we would like to avoid
``double-counting'' different anomalies with the same cause as different
bugs. It is a hard software engineering problem to establish the cause
of an anomaly/bug, which is out of scope of this paper.
To avoid this problem, however, we conservatively consider
only the top-most anomaly in each app in the top-10\%, as clearly,
anomalies in two different apps cannot have the same cause.

Through a manual inspection and triage by one of the authors, we
found 10 different types of possible bugs in our testing corpus
(Figure~\ref{fig:anomalies}), ranging from the benign to the
insidious. We have already seen instances of \#6 and \#10
(Figure~\ref{fig:ui-bug}) that could display corrupted or unclosable
dialog boxes. \#2 is serious as it could lead to an exception being
thrown due to a failed connection. \#5 would create a
crypto object that defaults to the semantically insecure ECB-encryption mode.
\#8 is serious, and could cause future attempts to open a socket, even by other
apps, to be blocked.

Figure~\ref{fig:results}(b)(i) shows the precision-recall plot for
these possible bugs in the top-10\% of anomaly scores. It can be seen that
at around the top 8\%, we reach full recall with 75\% precision or 25\%
false positive rate.  This is reasonable compared to industrial static
analysis tools such as Coverity that advocates a 20\% false positive rate for
``stable'' checkers~\cite{Bessey:2010:FBL:1646353.1646374}. Our method does not
rely on specified properties to check, and many of these bugs
cannot be easily expressed as a formal property for traditional static analyzers
to check.

After this threshold, the precision continues to drop,
and we conjecture that it will not increase any further, because almost all the
possible bugs have already been found.
To substantiate this conjecture, we would have to manually inspect
thousands of programs to qualitatively declare that all anomalies have been
triaged. Due to the practical infeasibility of this task, we instead
quantitatively {\em injected} anomalies into the remaining 90\% of
programs through mutations, and measured whether our model is
able to detect those mutations.
For each program, we mutated the API call before its accepting
states into one chosen randomly from $\Sigma$.

Figure~\ref{fig:results}(c) shows the anomaly scores before (dark) and
after (light) the mutation, and the cumulative mean of the relative
increase in the score (dashed line, secondary axis).
As a result of the mutation, the scores are greatly increased, sometimes
by 20 times or more, and the mean of the increase is about 4x.
That is, a mutation, on average, caused the anomaly score to increase by
4 times, indicating that our model detected the mutation.

Note that a random mutation, of course, has the possibility of {\em
reducing} the anomaly score of a program if it had a possible bug
and the mutation happened to fix it. However, since it is not very
likely for a random mutation to fix a possible bug, these instances were
few and far between.

\subsection{Role of Learning in Anomaly Detection}
\label{subsec:knn}

To evaluate the role of learning, we compared with a traditional outlier
detection method that does not require learning.
{\em k-nearest neighbor} (k-NN) outlier detection~\cite{altman1992knn} uses a distance
measure to compute the $k$ nearest neighbors of a given point within a
dataset. The larger the average distance to the k-NN, the more likely it
is that the point is an outlier, or anomaly. We already have a distance
measure between distributions: the KL-divergence between the behavior
model for the given program and a program in the corpus.

We implemented such a k-NN and compared our method with it by
conservatively setting $k = 1$. That is, the anomaly score of a given
program is the {\em smallest} KL-divergence with any program in
the corpus. However, even with this 1-NN anomaly score, a substantial
top 25\% of programs had a distance of infinity to the corpus,
thus providing no useful information about their anomalies.

The reason is that these programs simply happened to generate a behavior
that was not generated by {\em any} program in the corpus. This sets
$P_1(\theta)$ to a non-zero value and $P_2(\theta)$ to zero in the
KL-divergence formula (Equation~\ref{eqn:kld}) immediately making the
sum infinity. This is unreasonable because we clearly do not want to
call every behavior we have not observed in the training data an anomaly, but instead would like to
assign probabilities even to behaviors that were never seen before. That
is, we would like to {\em generalize} from the corpus. This is why
probabilistic specification learning is needed.

\subsection{Comparison with non-Bayesian methods}
\label{subsec:mining}

\begin{figure}
    \begin{tabular}{cc}
        \hspace{0.3cm}\includegraphics[scale=0.6]{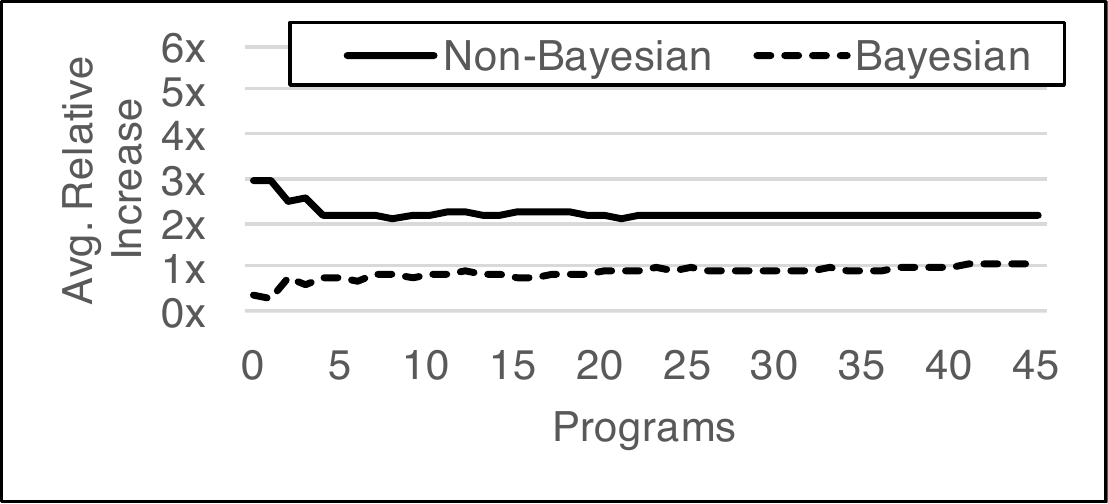} & (a) \\
        \hspace{0.3cm}\includegraphics[scale=0.6]{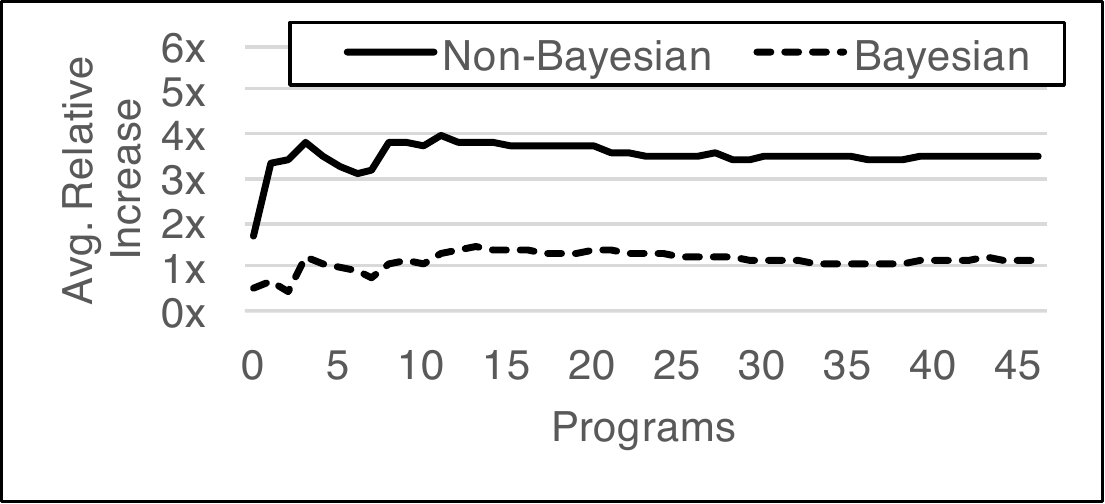} & (b)
    \end{tabular}
    \caption{Average relative increase in anomaly scores of ${\tt BluetoothSocket}$
        programs when the training corpus only uses the APIs (a) ${\tt
    BluetoothSocket}$,  ${\tt Cipher}$ (b) ${\tt AlertDialog.Builder}$,
${\tt BluetoothSocket}$, ${\tt Cipher}$}
    \label{fig:hetero}
\end{figure}

To see how the Bayesian framework helps in handling heterogeneity in the
corpus, we compared our method with a non-Bayesian specification learning method.
Existing state-of-the-art methods use n-grams~\cite{bugramASE16} or
RNNs~\cite{vechevPLDI14} to learn a (non-Bayesian) single probabilistic
specification of program behaviors. We implemented a non-Bayesian
specification learner as an RNN ({\em not} topic conditioned) and
trained it directly on the behaviors in our training corpus. We then
performed the same anomaly and bug detection experiment in
Section~\ref{subsec:anomaly}, querying the trained model with
behaviors in the testing program for inference. \ignore{ That is, we computed
the anomaly score as the KL-divergence between $P(\theta | \textbf{M})$
(the trained model) and $P(\theta|F)$ (the behavior model of program $F$),
with $X_F$ and $\Psi$ playing no role in the computation.}

Figure~\ref{fig:results}(b)(ii) shows the precision-recall rate for
the top-10\% of anomaly scores. Compared to our Bayesian method, the
non-Bayesian method performed poorly. Consider again a ``stable'' checker's
false-positive rate of 20\%, or 80\% precision. At this threshold (marked by
the red line), our Bayesian method has about 80\% recall compared to only
53\% for the non-Bayesian method. This shows that given a reasonable precision
threshold, our method is able to discover significantly more bugs compared
to the non-Bayesian method.
It is also worth noting that the non-Bayesian method
was unable to discover any possible bug that was not found by our
method.

\vspace{-0.6mm}
\paragraph{Effect of Heterogeneity.} We finally performed a series of
experiments by incrementally increasing the heterogeneity of the
training programs. First, as a baseline, we considered only
programs that use the ${\tt BluetoothSocket}$ API, and learned
from them both Bayesian and non-Bayesian specifications of their
behaviors. We then computed anomaly scores of the 45 testing programs that
use this API.

In the next step, we added to the training corpus programs that also use
the ${\tt Cipher}$ API, making the corpus more heterogeneous, and
learned new specifications. We then computed anomaly scores again,
but using the new learned specifications.
Figure~\ref{fig:hetero}(a) shows the average relative increase in
anomaly scores from using the old versus the new specifications.
Ideally, one would expect the scores to not change, because the addition
of programs that use the ${\tt Cipher}$ API---behaviors on which are
unrelated to the ${\tt BluetoothSocket}$ API---
should not have any effect on the scores. This is observed in the
Bayesian specification (dashed line), that lingers close to 1.0 on
average. However, the non-Bayesian specification (solid line) suffers
from about a 2x increase.

This was further evident when programs that also use the API ${\tt
AlertDialog.Builder}$ were considered for training, making the corpus
even more heterogeneous (this is the same training corpus in
Section~\ref{subsec:anomaly}). In Figure~\ref{fig:hetero}(b),
the relative increase in scores using the Bayesian specification is, on
average, close to 1.0, showing that it is robust to the increased
heterogeneity.  However, the non-Bayesian specification induces a 
further increase of about 3.5x in the scores.

We expect the gap to keep widening as more heterogeneous programs
are added to the corpus, at some point making the scores from the
non-Bayesian model meaningless. In contrast, the scores from our
Bayesian model would remain almost the same showing that the model is
able to ``focus'' on relevant parts of the learned specification, in
principle tolerating arbitrary heterogeneity.

\section{Related Work}\seclabel{relwork}

\paragraph{Learning Qualitative Specifications.}
The thesis that common patterns of execution can serve as a proxy for
specifications has been around since the early 2000s. Most efforts in
this
area~\cite{ABL02,Ammons03,Zhong09,LiFSE13,Ernst99,Nimmer02,Weimer05,Alur05,GouesTACAS09,Shoham07,Whaley02}
focus on qualitative specifications, typically finite automata. As
mentioned earlier, such qualitative specifications are problematic in
the presence of noise in the training data.


\paragraph{Learning Probabilistic Specifications.} There is also a
body of work
~\cite{ABL02,BeckmanPLDI11,Kremenek06,Livshits09,OcteauPOPL16,gveroPLDI13,vechevPLDI14}
that uses machine learning techniques to learn probabilistic
specifications from programs. Kremenek et
al.~\cite{Kremenek07,Kremenek06} use factor graphs constructed using
static analysis results to learn specifications on resource allocation
and release. {\sc Anek}~\cite{BeckmanPLDI11} uses 
annotations in APIs to infer specifications.  {\sc Merlin}~\cite{Livshits09} starts with a
given initial specification and refines it through factor graph
construction and inference.  Octeau et al.~\cite{OcteauPOPL16} use domain
knowledge to train probabilistic models of Android inter-component
communication.
{\sc JSNice}~\cite{raychev2015predicting} uses a
probabilistic graphical model to learn lexical and syntactical
properties of programs such as variable names and types for the
purpose of de-obfuscating Javascript programs. Some recent
efforts~\cite{nguyenFSE13,nguyenICSE15,bugramASE16} have also used {\em
n-gram} models to learn specifications on source code structure. {\sc
DeepAPI}~\cite{deepAPI} uses a neural encoder-decoder to learn
correlations between natural language annotations and API sequences.
{\sc Haggis}~\cite{idiomsFSE14} uses statistical techniques to learn
the structure of small code snippets (or ``idioms'') from a corpus.
The work in this space that is perhaps the closest to ours are two papers by
Raychev et al.~\cite{vechevPLDI14,vechevPOPL16}. These approaches
learn probabilistic models of program behavior from large code
corpora, using recurrent neural networks among other models.

The key difference between the above approaches and ours is
that these methods learn a {\em single} probabilistic
specification. In contrast, our approach learns a {\em family} of
probabilistic specifications simultaneously, and then specializes this
``big'' model to particular analysis tasks using Bayes' rule. As
demonstrated in our experiments, this hierarchical architecture is key
to tolerating heterogeneity.



In very recent work, Raychev et al.~\cite{vechevOOPSLA16}, 
also argue that having a single, universal
probabilistic model for code can be inadequate, and propose a
decision tree algorithm that is used to choose among a bag of
statistical models for tasks such as next-statement prediction.  While
philosophically aligned with our work, their efforts are quite different
in that while we argue for conditioning of models at the \emph{program
level}, they argue for conditioning of models at the \emph{statement
level} and focus their efforts on localized prediction tasks.  One could
imagine using a model similar to what they have proposed within our
framework as a replacement for our RNN-based $P(\theta|\Psi)$.
\ignore{, though
some technical difficulties would need to be overcome, such as the fact
that we require a generative model, while their model is
discriminative. }


\paragraph{Anomaly Detection.}
There is prior work on using learned models of executions in anomaly
detection~\cite{WasylkowskiFSE07,ECOOP10,hangal2002tracking,chilimbi2006heapmd,baah2006line,gao2007dmtracker,fu2009execution,bugramASE16}.
Aside from differing in the nature of specifications used, methods in
these categories tend to assign anomaly scores to individual behaviors
(generated statically or dynamically). While our method is able to
assign such scores, it is also able to produce
aggregate anomaly scores for programs.

\section{Conclusion}\seclabel{conc}

We have presented a Bayesian framework for learning
probabilistic specifications from large, heterogeneous code corpora,
and a method for finding likely software errors using this framework.
We have used an implementation of this framework, based on a
topic-model and a recurrent neural network, to detect API misuse in
Android, and shown that it can find multiple subtle bugs.

A key appeal of our framework is that it does not impose an a priori
limit on the size or heterogeneity of the corpus. In principle, our
training corpus could contain all the world's code, and it is our
vision to scale our method to settings close to this
ideal. Engineering instantiations of the framework that work at such
scale is a challenge for future work.

\ignore{
Another research direction is to study extensions of the approach to
problems such as program synthesis, repair, and optimization. A large
code corpus includes many insights about how humans implement, fix,
and optimize programs. One could imagine a statistical model that
learns such insights and serves them to algorithms that seek to
automate these tasks. The design and implementation of such models and
algorithms is an exciting research challenge.
}

\newpage
\def\bibfont{\footnotesize}
\bibliographystyle{authordate1}
\bibliography{refs}

\newpage
\appendix
\section{Constructing Program Models through Symbolic Execution}\seclabel{symexec}
Here, we define the process that transforms a given program into a
program model (Section~\ref{sec:formulation}) that our framework accepts
as input. We consider a simple programming language that
represents a program as a Control Flow Graph (CFG) $G$, where the
vertices are the locations in the program $\{l_1, \ldots, l_n\}$ and the
edges are basic program operations from the following types: (i)
assignments $x$\Assign$y$, (ii) assume statements {\sf assume($c$)} that
proceed with execution only if the Boolean condition $c$ is true, and
(iii) method invocations $m(a_1,\ldots,a_k)$ that call the method $m$
with arguments $a_1,\ldots,a_k$.

We construct the program model $F$ through a {\em symbolic execution} of
the CFG $G$. Symbolic execution~\cite{King76SE,tracercav12,KLEE}
executes a program with symbolic inputs, and keeps track of a {\em
symbolic state} that is analogous to a program's memory. A symbolic
state is a tuple $\langle l, v \rangle$ where $l$ is a program
location and $v$ is a symbolic store that is a mapping of variables to
symbolic values. The symbolic states encountered during symbolic
execution are exactly the states $\states$ in the program model.
The set of accepting states $Q_A$ typically consists of all states
$\langle l, \cdot \rangle$ at a particular location $l$ in the program.

The evaluation of a symbolic expression $e$ in a symbolic store $v$,
denoted as $\eval{e}{v}$, is defined as usual: $\eval{x}{v} =
v(x)$, $\eval{k}{v} = k$, $\eval{e_1 + e_2}{v} = \eval{e_1}{v} +
\eval{e_2}{v}$, $\eval{e_1 - e_2}{v} = \eval{e_1}{v} - \eval{e_2}{v}$ and so
on, where $x$ and $k$ are variables and constants, respectively. The
evaluation of a Boolean condition $\eval{c}{v}$ can also be defined in a
similar manner. Note that $\eval{c}{v}$ requires a theorem prover to
evaluate, and the completeness of this theorem prover affects the
exclusion of infeasible states from the program model.
\ignore{, which we assume is sound but not necessarily complete. That
is, it should decide that $\eval{c}{v}$ is $false$ only if it is indeed
so.}

Symbolic execution begins with an initial state $q_0 = \langle l_0,
\emptyset \rangle$ where $l_0$ is the (special) initial location
of the program with an empty store. Then, recursively, given a state $q
= \langle l, v \rangle$, symbolic execution proceeds as follows:

\begin{itemize}
\item[-] If $G$ has an edge from $l$ to $l'$ where the operation is
    $x$\Assign$y$, then $q' = \langle l', v[x \mapsto \eval{y}{v}]
    \rangle$ is added to $\states$ and $\transition{q}{q'}{\epsilon}{1}$
    is added to $\delta$, where $\epsilon$ is the empty symbol in
    $\Sigma$.

\item[-] If $G$ has edges from $l$ to $l_1',\ldots, l_m'$
    with operations {\sf assume($c_1$)}, $\ldots$, {\sf assume($c_m$)}
    respectively, then each condition is first evaluated for
    satisfiability: let $C$ be the set $\{l_j' ~|~ \eval{c_j}{v} =
    true\}$. Then, for each $l_j'$ in $C$, a state $q_j' = \langle l_j',
    v \rangle$ is added to $\states$ and
    $\transition{l}{l_j'}{\epsilon}{1/|C|}$ is added to $\delta$. By
    default, we give a uniform probability of $1/|C|$ to all feasible
    branches, but this can be enhanced if the distribution of inputs to
    the program is known, or through external techniques such as {\em
    model counting}.

\item[-] If $G$ has an edge from $l$ to $l'$ with the operation
    $m(a_1,\ldots,a_k)$, then a state $q' = \langle l', v \rangle$ is
    added to $\states$, and a transition $\transition{q}{q'}{m}{1}$ is added to
    $\delta$ if $m$ is in the alphabet $\Sigma$, otherwise,
    $\transition{q}{q'}{\epsilon}{1}$ is added to $\delta$.
\end{itemize}

\section{Counteracting Bias in Anomaly Scores}\seclabel{bias}
\label{subsec:bias}

As mentioned before, the estimator used to compute anomaly scores is
biased because the logarithm and expectation operators do not commute. 
However, we can develop an algorithm
to counter-act the bias in estimating the bias of a sampling-based estimator for a logarithm
using a Taylor series expansion.  Assume that we have an estimator
of the form $X = \sum_{s \in S} f(s)$ for some sample set $S$ and function $f(s)$; 
there are two such estimators in the 
computation of the anomaly score algorithm of Section 4.4.
Also let $\hat{p}$ denote $E[X]$.
Expanding the Taylor series for $\log(X)$ 
around the point $\hat{p}$ 
we have $\log(X) =$
$$\log(\hat{p}) + \frac{(X - \hat{p})}{\hat{p}} - 
	\frac{(X - \hat{p})^2}{2\hat{p}^2} +
	\frac{(X - \hat{p})^3}{3\hat{p}^3} - 
	\frac{(X - \hat{p})^4}{4\hat{p}^4} + ...$$
So, $E[\log(X)] =$ 
$$\log(\hat{p}) + \frac{E[(X - \hat{p})]}{\hat{p}} - 
        \frac{E[(X - \hat{p})^2]}{2\hat{p}^2} +
        \frac{E[(X - \hat{p})^3]}{3\hat{p}^3} - ...$$
Then the bias is $E[\log(X)] - \log(E[X])$, or
$$\frac{E[(X - \hat{p})]}{\hat{p}} - 
        \frac{E[(X - \hat{p})^2]}{2\hat{p}^2} +
        \frac{E[(X - \hat{p})^3]}{3\hat{p}^3} - ...$$
The numerator in the $i$th term in this series corresponds to the
$i$th central moment of $X$.  Since $X$ is computed as an average
of a number of independent, identically distributed samples, according
to the Central Limit Theorem, $X$ is asymptotically normally distributed.
Given this, and using the central moments of the normal distribution,
we can write the bias as:
$$-\frac{\sigma^2(X)}{2\hat{p}^2} -\frac{3\sigma^4(X)}{4\hat{p}^4} -
	\frac{15\sigma^6(X)}{6\hat{p}^6} - 
	\frac{105\sigma^8(X)}{8\hat{p}^8} - ...$$
In practice, an estimate that takes into account the first three or four terms
should be adequate.
When estimating those terms,
$\hat{p}$ can be estimated
using the observed value of 
$X$; that is, $\hat{p} \approx
\sum_{s \in S} f(s)$.
$\sigma^2(X)$ is the variance of $X$ and can be estimated in many ways;
we advocate using a
bootstrap resampling procedure \cite{efron1994introduction}.  
That is, we uniformly
resample (with replacement) a set $S'$ of $|S|$ items from $S$, and use this
to produce a simulated value $\sum_{s \in S'} f(s)$ for $\hat{p}$.
We perform this
resampling procedure many times, and use the observed variance as an estimate
for $\sigma^2(X)$. In this way, it is possible
to obtain an estimate for $bias(\theta, \Pi'_F, \psi_F)$.
Plugging this into the equation at the end of Section 4.4 then gives us
our final estimator.

One last issue to consider is how to 
to compute the variance of this estimator so that we can know the
error of our estimate and stop sampling at the appropriate
time.  We also advocate a bootstrap resampling procedure for this task.
We simply re-sample $m$ $(\theta, \Pi'_F, \psi_F)$
triples
from the set of sampled $(\theta, \Pi'_F, \psi_F)$ triples.
We then compute Section 4.4's estimator for each such triple.  Taking the average
value over all of the triples produces a simulated estimate 
for the anomaly score.
Repeating this process many times and computing the 
empirically observed variance produces an estimate for the variance of the anomaly score.

\section{Gibbs Sampling to Approximate $P(\Psi | X; \textbf{M})$}\seclabel{gibbs}
LDA models a document as a distribution over topics, and a topic as a
distribution over words in the vocabulary. An LDA model is thus fully
characterized by the following variables: (i) $\alpha$ and $\eta$,
hyper-parameters of a Dirichlet prior that chooses the topic
distribution of each document and  the word distribution of each topic,
respectively (ii) $\Psi_{F_i}$, the topic distribution of document
$X_{F_i}$, (iii) $\beta_k$, the word distribution of topic $t_k$, (iv)
$z_{i,j}$, the topic that generated word $j$ in document $i$, and (v)
$w_{i,j}$, the $j$th word in document $i$.
With these parameters, the generative process of documents, according to
LDA, is as follows:

\begin{itemize}
    \item[1.] Decide on a topic distribution for document $X_{F_i}$:
        $\Psi_{F_i} \sim \Dir_K(\alpha)$, where $\alpha$ is a $K$-length
        vector of reals. This sampling would produce a distribution over
        $K$ topics.
\item[2.] Decide on a word distribution for topic $t_k$: $\beta_k \sim
    \Dir_{|\Sigma|}(\eta)$.
\item[3.] For each word position $i, j$, $1 \leq i \leq M$ and $1 \leq j \leq N_i$,
    \begin{itemize}
    \item[(a)] Decide on a topic for the word to be generated at $i, j$:
        $z_{i,j} \sim \Categorical_K(\Psi_{F_i})$.
    \item[(b)] Generate the word: $w_{i,j} \sim
        \Categorical_{|\Sigma|}(\beta_{z_{i,j}})$.
    \end{itemize}
\end{itemize}

\noindent
The total probability of all variables in the model is then
$$\prod_{k=1}^K P(\beta_k | \eta) \prod_{i=1}^M P(\Psi_{F_i} | \alpha)
\prod_{j=1}^N P(z_{i,j} | \Psi_{F_i}) P(w_{i,j} | \beta_{z_{i,j}})
$$

We refer the reader to~\cite{LDA} for details on training an LDA model.
The result of training is a learned value for all latent variables in
the model, particularly all the $\beta_k$. During inference, we are
given a particular $X_F$ and would like to compute the posterior
distribution $P(\Psi | X_F ; \textbf{M})$.

In LDA terms, $\Psi$ would be a topic distribution for $X_F$
(we drop the $i$ subscript since we are now only working with one
document). In order to estimate this distribution, we employ a standard
Markov Chain Monte Carlo (MCMC) method called Gibbs sampling.
First, we sample $\Psi$ from a Dirichlet prior parameterized by some
randomly chosen $d$: $\Psi \sim \Dir_K(d)$. Then, each step of Gibbs sampling
re-computes $\Psi$ using Bayes' rule:

\begin{itemize}
\item[1.] For every word $w_j$ in $X_F$, decide on a topic $t_k$ that
could have generated $w_j$ by sampling from the distribution $$P(t_k |
w_j) = \frac{P(w_j | t_k) P(t_k)}{P(w_j)}$$ where $P(w_j | t_k)$ is
the probability of the word $w_j$ according the learned model $\beta_k$,
$P(t_k)$ is obtained from the current value of $\Psi$ and $P(w_j) =
\sum_{k=1}^K P(w_j | t_k) P(t_k)$ is a normalization term.

\item[2.] Compute the vector $n = (n_1, \ldots, n_K)$ where $n_k$ is the
number of words in $X_F$ that were generated by topic $t_k$.

\item[3.] Re-sample $\Psi$ from the distribution $\Psi \sim \Dir_K(d + n)$.
\end{itemize}

Gibbs sampling guarantees that by repeating these steps for some number
of iterations (that depends on the value of $d$), the samples of $\Psi$
would eventually converge to coming from the posterior
distribution $P(\Psi | X_F; \textbf{M})$, thus approximating this
distribution through sampling.

\end{document}